\documentclass[aps,prc,amsmath,amssymb,nofootinbib,%
twocolumn,superscriptaddress,nolongbibliography]{revtex4-2}

\usepackage{amsfonts,amsmath,amssymb}
\usepackage{mathrsfs}
\usepackage{graphicx}
\usepackage{multirow}
\usepackage{epsfig}
\usepackage{graphicx}
\usepackage{subfigure}
\usepackage{booktabs}
\usepackage{array}
\usepackage{tabularx}
\usepackage{braket}
\usepackage{bm}
\usepackage{diagbox}
\usepackage{verbatim}

\usepackage[usenames]{color}
\usepackage[dvipsnames]{xcolor}
\usepackage[utf8]{inputenc}
\usepackage{CJK}
\usepackage{hyperref}
\hypersetup{
  colorlinks=true,        
  linkcolor=ForestGreen,    
}

\renewcommand{\vec}[1]{\mathbf{#1}}

\newcommand{\chn}[3]{{{}^{#1}\!{#2}_{#3}}}
\newcommand{\cs}[2]{\chn{#1}{S}{#2}}

\newcommand{\cd}[2]{\chn{#1}{D}{#2}}

\newcommand{\csd}{{\cs{3}{1}-\cd{3}{1}}}

\newcommand{\NNLO}{N$^2$LO}

\newcommand{\dd}{\mathrm{d}}

\begin{document}

\title{Three-nucleon contact forces in the Jacobi partial-wave basis}

\author{Lin Zuo}

\author{Hao Yang}
\email{yanghao2023@scu.edu.cn}
\affiliation{College of Physics, Sichuan University, Chengdu 610065, Sichuan Province, China}

\author{Bingwei Long}
\email{bingwei@scu.edu.cn}
\affiliation{College of Physics, Sichuan University, Chengdu 610065, Sichuan Province, China}

\affiliation{Southern Center for Nuclear-Science Theory (SCNT), Institute of Modern Physics, Chinese Academy of Sciences, Huizhou 516000, Guangdong Province, China}

\date{October 18, 2025}
    
\begin{abstract}
We construct the three-nucleon contact potentials in the Jacobi partial-wave basis. The potentials are built in the separable form as the products of the antisymmetrized three-nucleon states in which the nucleons are arbitrarily close to each other. We compile the three-nucleon contact potentials up to $\mathcal{O}(Q^2)$. These contact potentials are by construction independent operators under exchanges of any nucleon pairs.  
\end{abstract}
	
\maketitle

\section{Introduction}
Three-nucleon forces have gained increasing attention in the recent development of nuclear interactions, especially within the framework of effective field theory. 
In chiral effective field theory (EFT), the first non-vanishing three-nucleon ($3N$) potential appears at next-to-next-to-leading order (\NNLO) in Weinberg's power counting~\cite{Weinberg:1990rz, Weinberg:1991um}. 
The development of $3N$ potentials in chiral EFT has focused mainly on the derivation of pion-exchange forces~\cite{vanKolck:1994yi, Epelbaum:2002vt, Bernard:2007sp, Bernard:2011zr, Krebs:2012yv, Krebs:2013kha}. 
However, there are evidences that power counting of $3N$ forces may need adjustments\cite{Hammer:2019poc, Kievsky:2016kzb}. 
$3N$ forces are much more important in Pionless EFT. 
There is one already at the leading order (LO) ~\cite{Bedaque:1997qi, vanKolck:1998bw, Bedaque:1998mb}, which is the momentum-independent $3N$ force crucial for producing the triton. 
Because pions are integrated out in Pionless EFT, all interactions are zero-ranged, i.e., both nucleon-nucleon ($NN$) and higher-body potentials are contact operators.

In this paper, we focus on the $3N$ contact interactions. 
Aside from the usual symmetry requirements such as rotational, time-reversal, isospin invariance etc., a nontrivial aspect in constructing these interactions is to identify an independent set of operators that remain invariant under exchanges of any two nucleons.
For instance, one can construct six momentum-independent $3N$ contact operators through combinations of spin and isospin Pauli matrices, expressed in momentum space as follows\cite{Hebeler:2020ocj}:
\begin{align}
    V_\text{3CT} &= \sum_{i\neq j\neq k}\Big{[}\beta_1 +\beta_2\boldsymbol{\sigma}_i \cdot \boldsymbol{\sigma}_j+\beta_3\boldsymbol{\tau}_i \cdot \boldsymbol{\tau}_j\nonumber\\
    & \quad +\beta_4\boldsymbol{\sigma}_i \cdot \boldsymbol{\sigma}_j\boldsymbol{\tau}_i \cdot \boldsymbol{\tau}_j+ \beta_5\boldsymbol{\sigma}_i \cdot \boldsymbol{\sigma}_j\boldsymbol{\tau}_j \cdot \boldsymbol{\tau}_k\nonumber\\
    &\quad + \beta_6 \big ( (\boldsymbol{\sigma}_i\times \boldsymbol{\sigma}_j) \cdot \boldsymbol{\sigma}_k\big ) \big ( (\boldsymbol{\tau}_i\times \boldsymbol{\tau}_j )\cdot \boldsymbol{\tau}_k \big )\Big{]}\, ,\label{eqn:Leading3NF}
\end{align}
where $i, j, k$ are the indexes of the nucleons. 
However, once sandwiched between antisymmetrized $3N$ states, they are found to be correlated to one another. Therefore, we only need to consider one of them, or any combination of them by our choice\cite{Epelbaum:2002vt}. 

For the momentum-dependent operators, the complexity increases rapidly because there are four momenta involved in the construction: two in the initial and two in the final states, if we remove the center-of-mass motion. 
Our goal is to determine a set of independent $3N$ contact operators with two powers of momenta, referred to as $\mathcal{O}(Q^2)$ operators in the literature of chiral EFT where $Q$ denotes generically all the external momenta.

A novel method is presented in the paper to tackle this problem.
We start with the consideration that contact interactions can be written in the separable form, more specifically, the outer product of two $3N$ states describing three nucleons arbitrarily close to each other,
referred to as ``zero-range'' states in the paper:
\begin{equation}
    W \propto \ket{H_c} \bra{H_d} + h.c.
\end{equation}
Requirement of antisymmetrization imposes another important constraint on 
$\ket{H_{c, d}}$. With these two constraints, we are able to establish a finite-dimension eigenvector equation in the Jacobi partial-wave basis, of which $\ket{H_{c, d}}$ are the solutions.

The same problem was tackled in Refs.~\cite{Girlanda:2011fh, Girlanda:2020pqn, Witala:2022rzl, Nasoni:2023adf},
where the $3N$ contact Lagrangian terms were first enumerated as polynomials of the nucleon fields and the Fierz transformations were then applied to check the correlations among the listed operators, so as to extract an independent set. 
The methodology of these works is quite different from ours.

The paper is organized as follows. 
In Sec.~\ref{sec:NNpart}, we demonstrate our method with the $NN$ contact interactions.
In Sec.~\ref{sec:3NZeroRangeState}, we show how to construct the antisymmetrized, zero-range $3N$ states.
This is followed by building the $3N$ contact potentials up to $\mathcal{O}(Q^2)$ in Sec.~\ref{sec:CntOpt}.
We discuss the transformation from the plane-wave basis to the momentum-space partial-wave basis in Sec.~\ref{sec:plane}.
Finally a summary is offered in Sec.~\ref{sec:conclusion}.

\section{\texorpdfstring{$NN$}{NN} contact interactions~\label{sec:NNpart}}

It is instructive to show how our approach works in the two-body sector where the formulation of the contact interactions has been long established.  
For nonrelativistic $NN$ dynamics, any contact interaction can be cast in a generic separable form:
\begin{equation}
    V_\text{CT} = C \left(\ket{g_c} \bra{g_d} + h.c. \right)\, , 
    \label{eqn:Sep2NF}
\end{equation}
where $C$ is the coupling constant. The states $\ket{g_c}$ and $\ket{g_d}$ are zero-range $NN$ wave functions where the two nucleons have zero separation, to the extent allowed by the ultraviolet regulator of the EFT. $\ket{g_c} \bra{g_d}$ is then interpreted as the transition from one zero-range state $\ket{g_d}$ to another $\ket{g_c}$. In a field-theoretic approach, one uses  Lagrangian terms to describe $NN$ contact interactions, typically quartic operators in nonrelativistic nucleon fields, from which one can derive four-nucleon vertexes that conform to Eq.~\eqref{eqn:Sep2NF}. This procedure can be found in the literature of chiral EFT and Pionless EFT, for instance, in Refs.~\cite{Kaplan:1996xu, Beane:1997pk, Cohen:1996my, Scaldeferri:1996nx, Gegelia:1998gn, Ordonez:1993tn, vanKolck:1998bw, Dietz:2021haj, vanKolck:1999mw, vanKolck:1994yi}
 
Due to the indistinguishability of the nucleons, $V_\text{CT}$ must be invariant under the interchange of the two nucleons:
\begin{equation}
    P_{12}^{-1}\, V_\text{CT} \, P_{12} = V_\text{CT} \, .
    \label{eqn:PermInv}
\end{equation}
Or equivalently, the zero-range $NN$ state must be antisymmetric:
\begin{equation}
    P_{12} |g \rangle = - |g \rangle \, ,
\end{equation}
where we have dropped the subscripts of $\ket{g_{c(d)}}$ to simplify the notation. 
The $NN$ partial-wave basis is automatically antisymmetrized, as long as the sum of the orbital angular momentum $l$, spin $s$ and isospin $t$ of the $NN$ pair is odd, i.e., $l+s+t = \rm odd$:
\begin{align}
    P_{12} | l s t \rangle &= (-1)^{l + s + t} | l s t \rangle= - \ket{l s t}, \label{eqn:EigenP12}
\end{align}
Therefore we will expand $\ket{g}$ in the partial-wave basis.

The other crucial constraint on $\ket{g}$ is that $\ket{g}$ is a zero-range $NN$ state.
In coordinate space, its wave function is a product of $\delta$ functions smeared by the ultraviolet regulator: 
\begin{equation}
    \langle r, l s t|g \rangle \propto \delta_\Lambda (r)\, ,\label{eqn:giWFs}
\end{equation}
where $r$ is the relative distance between the two nucleons,
and the spread of $\delta_\Lambda(r)$ is regulated by the inverse momentum cutoff $\Lambda^{-1}$. As long as $\Lambda$ is sufficiently large, the momentum-space wave function of $\ket{g}$ has the following form:
\begin{align}
    \langle p, l s t | g_n, lst\rangle  &\propto \int dr r^2 \,j_{l}(p r) \, \delta_\Lambda (r) \propto p^l p^{2n}
    \, ,\label{eqn:gAsmpt}
\end{align}
where $j_{l}(x)$ is the spherical Bessel function of the first kind. By choosing different values taken by $n$, we can construct many variants of $\ket{g}$ with increasing powers of momentum $p$ up to the desired EFT order.

As an example, we build the $SD$ mixing contact interactions for the coupled channel of $\csd$. 
The zero-range states in $\cs{3}{1}$ and $\cd{3}{1}$:
\begin{align}
    \braket{p, \cs{3}{1} | g_0\, \cs{3}{1}} &= 1\, , \\
    \braket{p, \cs{3}{1} | g_1\, \cs{3}{1}} &= p^2\ , \\
    \braket{p, \cd{3}{1} | g_0\, \cd{3}{1}} &= p^2\, , \\
    \braket{p, \cs{3}{1} | g_1\, \cd{3}{1}} &= p^4\ , \\
    \cdots \quad & \nonumber   
\end{align}
The $SD$ mixing contact interactions are then assembled as the outer product of $\ket{g, lstj}$ where $j$ is the total angular momentum, sorted below by their powers in momenta:
\begin{align}
    V_{SD}^{(2)} &= C_{2} \left( \ket{g_0 \cd{3}{1}}\bra{g_0 \cs{3}{1}} + h.c. \right)\, , \\
V_{SD}^{(4)} &= C_{4} \left( \ket{g_0 \cd{3}{1}}\bra{g_1 \cs{3}{1}} + h.c. \right) \nonumber \\
& \quad + C_{4}^{\prime} \left( \ket{g_1 \cd{3}{1}}\bra{g_0 \cs{3}{1}} + h.c. \right) \, ,\\
& \cdots \nonumber
\end{align}
and sandwiching these operators between the partial-wave basis yields
\begin{align}
    \braket{p', \cd{3}{1} |V_{SD}^{(2)}| p, \cs{3}{1}} &= C_{2} {p'}^2 \, ,\\
    \braket{p', \cd{3}{1} |V_{SD}^{(4)}| p, \cs{3}{1}} &= C_{4} {p'}^2 \left( {p'}^2 + p^2 \right) + C_4^\prime {p'}^4 \, , \\
& \cdots \nonumber
\end{align}

Further considerations can be made, including restoring order-by-order Lorentz invariance for small momenta and eliminating the redundancy due to the nucleon equation of motion. 
All these may reduce the number of independent contact interactions. 
In fact, one can show that the $C_4$ term and $C'_4$ term can be related by the equation of motion~\cite{Beane:2000fi}. 
But we will not account for these constraints in the present paper.

\section{Zero-range three-nucleon states
\label{sec:3NZeroRangeState}}

The idea remains much unchanged as we turn to the $3N$ system. A separable $3N$ contact potential is built from the product of antisymmetrized, zero-range $3N$ states:
\begin{equation}
    W_\text{CT} \propto \ket{H_c} \bra{H_d} + h.c. \label{eqn:OutPrdtHiHj}    
\end{equation}
Here $\ket{H_{c(d)}}$ flips its sign under the exchange of any two nucleons:
\begin{equation}
    P_{kl} \ket{H} = - \ket{H}\, ,
\end{equation}
where $P_{kl}$ exchanges the two nucleons labeled by $k$ and $l$. We also drop the subscript of $\ket{H_{c(d)}}$ since this is a universal property.

We find it convenient to express $\ket{H}$ in the Jacobi partial-wave basis and, in particular, using the \emph{uncoupled} $LS$ scheme where $L$ is the total orbital angular momentum and $S$ the total spin:
\begin{equation}
    \ket{pq\alpha} \equiv \ket{
    pq\; (l\lambda) {L} M_L\, \left (s \frac{1}{2}\right )  {S} M_S \, \left (t \frac{1}{2}\right ) T M_T}
    \, ,\label{complete_basis} 
\end{equation}
where $p$ is the relative momentum between the pair of nucleon $1$ and $2$, $q$ the momentum of the third nucleon--- the spectator--- relative to the center of mass of the pair.
Here $\alpha$ is a collective label to denote all the discrete quantum numbers of a given channel, including $lsjt$ the previously defined quantum numbers of the pair, $T$ the total isospin of the $3N$ system, 
and $\lambda$ the orbital angular momentum of the spectator. $M_{L(S)}$ is the $z$ component of the total orbital (spin) angular momentum and $M_T$ the third component of the total isospin.
The parity $\mathcal{P}$ is determined by $l+\lambda$:
\begin{align}
    \mathcal{P} &= (-1)^{l+\lambda} \, . 
\end{align}
    
The three-body cyclic permutation is the key in our approach~\cite{Glockle-Fewbody83}:
\begin{align}
    P_{123} = P_{12}P_{23}\, ,\; P_{123}^{-1} = P_{13}P_{23} \, .
    \label{eqn:P123Def}
\end{align}
From the above definitions, one can derive the following identity:
\begin{equation}
    P_{123}^2 = P_{123}^{-1}\, ,
    \label{eqn:CyclicAnti}
\end{equation}
and express $P_{23}$ and $P_{13}$ in terms of $P_{12}$ and $P_{123}$:
\begin{align}
    P_{23} &= P_{12} P_{123} \, ,\\
    P_{13} &= P_{123} P_{12} \, .
\end{align}
By construction, the subsystem of nucleon $1$ and $2$ is already antisymmetrized in the Jacobi partial-wave basis:
\begin{align}
    P_{12} \ket{pq\alpha} &= -\ket{pq\alpha} \, .
    \label{eqn:P12Anti}   
\end{align}
To ensure the total antisymmetrization of $\ket{H}$, i.e.,
\begin{align}
    P_{23} \ket{pq\alpha} &= -\ket{pq\alpha} \, , \label{eqn:P23Anti}\\
    P_{13} \ket{pq\alpha} &= -\ket{pq\alpha} \, ,
    \label{eqn:P13Anti}
\end{align}
we only need to impose one additional requirement that $\ket{H}$ remain invariant under the cyclic permutation $P_{123}$:
\begin{equation}
    P_{123}|H \rangle = |H \rangle\, . \label{eqn:HPInv}
\end{equation}

We will also need to utilize the zero-ranged nature of $\ket{H}$. 
It will allow $\braket{pq\alpha|H}$ and its permuted wave function $\braket{pq\alpha|P_{123}|H}$ to be expanded over a finite-dimension subspace of $\{\alpha\}$, for a given momentum power $Q^\nu$ where $Q$ represents either $p$ or $q$.
As a result, solving Eq.~\eqref{eqn:HPInv} for $\ket{H}$ reduces to solving for the eigenvectors of $P_{123}$ with the eigenvalue of $1$ in a finite-dimensional space.

We begin to establish such eigenvector equations by reviewing the representation of $P_{123}$ in the basis of $\ket{pq\alpha}$~\cite{Glockle:1996jg}:
\begin{align}
    &\bra{p'q' \alpha'}P_{123}\ket{pq \alpha} 
    = \delta_{L' L} \delta_{S' S} \delta_{T' T}
    \delta_{M_L' M_L} \delta_{M_S' M_S} \delta_{M_T' M_T} \nonumber\\
    & \quad \times \int_{-1}^1 dx \dfrac{\delta(p-\Pi_1)}{p^{l+2}} \dfrac{\delta(q-\Pi_2)}{q^{\lambda+2}} G_{\alpha' \alpha}(p'q'x) \, , \label{eqn:RepP123}
\end{align}
where
\begin{align}
    \Pi_1 &= \sqrt{ \dfrac{1}{4}p'^2 + \dfrac{9}{16}q'^2 + \dfrac{3}{4}p'q'x }, \\ \nonumber
    \Pi_2 &= \sqrt{ p'^2 + \dfrac{1}{4}q'^2 - p'q'x}\, ,
\end{align}
\begin{align}
    G_{\alpha '\alpha}(p'q'x) &= \sum_{k} P_k(x)
    \sum_{\substack{l_1+l_2=l \\ 
    \lambda_1+\lambda_2=\lambda}}
    (p')^{l_1+\lambda_1}\nonumber\\
    &\quad \times (q')^{l_2+\lambda_2}\tilde{g}_{\alpha '\alpha}^{kl_1l_2\lambda_1\lambda_2}\, ,\label{eqn:Gpqx}
\end{align}
and all the primed quantum numbers such as $l', \lambda', s', \cdots$, collectively denoted by $\alpha'$, identify the final state.
The coefficients $\tilde{g}_{\alpha '\alpha}^{kl_1l_2\lambda_1\lambda_2}$ are defined as follows:
\begin{align}
    &\tilde{g}_{\alpha '\alpha}^{kl_1l_2\lambda_1\lambda_2}\nonumber \\
    &=(-1)\sqrt{\hat{s'}\hat{t'}\hat{l'}\hat{\lambda'}\hat{l}\hat{\lambda}\hat{s}\hat{t}} 
    \left \{ 
    \begin{array}{ccc}
        \frac{1}{2} &\frac{1}{2} & s'\\[0.5ex] \frac{1}{2}&  {S}&s \end{array} \right \}  \left \{ \begin{array}{ccc}  \frac{1}{2} &\frac{1}{2} & t'\\[0.5ex] \frac{1}{2}&  {T}&t 
        \end{array}
    \right \} \nonumber\\
    &\quad \times (-1)^{ {L}}(-1)^{k+\lambda+\lambda_1}{\left (\frac{1}{2}\right )}^{l_1+\lambda_2+1}\left (\dfrac{3}{4} \right)^{l_2}\nonumber\\
    &\quad \times\sqrt{\dfrac{(2l+1)!}{(2l_1)!(2l_2)!}}\sqrt{\dfrac{(2\lambda+1)!}{(2\lambda_1)!(2\lambda_2)!}}\sum_{f f'} \sqrt{\hat{f}\hat{f'}}\nonumber\\
    &\quad \times C_{l_1 \lambda_1 f}^{0\,0\,0} C_{l_2 \lambda_2 f'}^{0\,0\,0}
    \left \{ 
        \begin{array}{ccc}  l_1 &l_2 & l\\ \lambda_1& \lambda_2&\lambda \\ f &f' &  {L}
        \end{array}
    \right \}\nonumber\\
    &\quad \times\left \{ \begin{array}{ccc} l'&f&k\\ f'&\lambda'& {L} \end{array} \right \} C_{l' f \,k}^{0\,0\,0}C_{\lambda'f' k}^{0\,0\,0}\, . 
    \label{eqn:Smallg}
\end{align} 
One of the advantages offered by the Jacobi partial-wave basis is that $P_{123}$ conserves each of $L$, $S$, $T$ and parity $\mathcal{P}$, as evidenced by the Kronecker symbols in Eq.~\eqref{eqn:RepP123}. 
This implies that a permutation-invariant state $\ket{H}$ has definite values of 
$LST\mathcal{P}$.

Although $P_{123}$ and $P_{123}^{-1}$ are generally distinct operators, their matrices
in the basis of $|pq\alpha\rangle$ are identical~\cite{Glockle-Fewbody83}, as the Jacobi partial-wave basis inherently antisymmetrizes the pair of nucleon $1$ and $2$:
\begin{equation}
    \bra{p'q'\alpha'} P_{123} \ket{pq\alpha} = \bra{p'q'\alpha'} P_{123}^{-1} \ket{pq\alpha} \, .
\end{equation}
Furthermore, $P_{123}$ is symmetric in the basis of $|pq\alpha\rangle$:
\begin{equation}
    \bra{p'q'\alpha'} P_{123} \ket{pq\alpha} = \bra{pq\alpha} P_{123} \ket{p'q'\alpha'} \, .
\end{equation}

Just as shown in the $NN$ case by Eq.~\eqref{eqn:gAsmpt}, the momentum-space wave function of $\ket{H}$ as a zero-range state is constrained by the orbital angular momenta $l$ and $\lambda$:
\begin{equation}
    \langle pq\alpha \ket{H} = \psi_\alpha p^l q^{\lambda} p^{2m} q^{2n} \, ,\label{eqn:Def_H}    
\end{equation}
where $\psi_\alpha$ are the coefficients to be determined by Eq.~\eqref{eqn:HPInv}.
We classify $\ket{H}$ according to the power of momenta in its wave function:
\begin{equation}
    \nu \equiv l + \lambda + 2(m + n)\, .
    \label{eqn:DefNu}
\end{equation}
To build $3N$ contact interactions of $\mathcal{O}(Q^2)$, we need to consider $\nu = l+\lambda = 0$, $1$, or $2$.

As we will see, when $m = n = 0$, the value of $L$ will be greatly restricted for a given $\nu$. 
Therefore, we will focus on $m = n = 0$ before considering the cases where they do not vanish. 

\subsection{\texorpdfstring{$m = n = 0$}{mn0}\label{sec:mn0}}

Whether the three particles described by $\ket{H}$ are close to one another is independent of how each particle is labeled by their indices; therefore, the permuted state $P_{123} \ket{H}$ is zero-range too.
Consequently, one expects
\begin{equation}
    \bra{p'q'\alpha'}P_{123}\ket{H} \propto {p'}^{l'} {q'}^{\lambda'}\, .
    \label{eqn:Local_PH}
\end{equation} 
Because the scaling property of $\ket{H}$ remains unchanged after any exchange of particles, $l' + \lambda'$ of the permuted state $P_{123}\ket{H}$ has the definite value $\nu$:
\begin{equation}
    l' + \lambda' = \nu \, .\label{eqn:ScalTrans}
\end{equation}

We will prove that in order to satisfy Eqs.~\eqref{eqn:Local_PH} and \eqref{eqn:ScalTrans}
for a given $\nu$, the channels for expanding $\ket{H}$ are subject to the following conditions:
\begin{gather}
    l + \lambda = \nu \, ,\label{eqn:llmbdeqnu} \\
    L = \nu \quad\text{or}\quad \nu - 1 \quad \text{if} \;\; \nu \geqslant 2\, . \label{eqn:Leqnu}
\end{gather}
We use Eq.~\eqref{eqn:RepP123} and Eq.~\eqref{eqn:Def_H} to express the permutation-transformed wave function of $\ket{H}$:
\begin{align}
&\braket{ p'q' \alpha'|P_{123}|H}\nonumber \\
     &= 
2 \sum_{
    \substack{
    \alpha,\, l_1+l_2=l \\ 
    \lambda_1+\lambda_2=\lambda}} \psi_{\alpha}\, (p')^{l_1+\lambda_1} (q')^{l_2+\lambda_2} 
  \tilde{g}_{\alpha' \alpha}^{0l_1 l_2 \lambda_1 \lambda_2}\, ,
    \label{eqn:pqaPHkEqs0}        
\end{align}
where we note that only $k = 0$ survives the integration over the Legendre polynomials in Eq.~\eqref{eqn:RepP123} and that $\tilde{g}_{\alpha' \alpha}^{0 l_1 l_2 \lambda_1 \lambda_2}$ has a simplified expression for $k=0$:
\begin{align}
     \tilde{g}_{\alpha' \alpha}^{0l_1 l_2 \lambda_1 \lambda_2}&=   
(-1)\sqrt{\hat{s'}\hat{t'}\hat{l}\hat{\lambda}\hat{s}\hat{t}} \left \{ 
    \begin{array}{ccc}
        \frac{1}{2} &\frac{1}{2} & s'\\ [0.5ex]
        \frac{1}{2}&  {S}&s \end{array} \right \}  \left \{ \begin{array}{ccc}  \frac{1}{2} &\frac{1}{2} & t'\\ [0.5ex]
        \frac{1}{2}&  {T}&t 
        \end{array}
    \right \} \nonumber\\
    &\quad \times (-1)^{\lambda+\lambda_1}{\left (\frac{1}{2}\right )}^{l_1+\lambda_2+1}\left (\dfrac{3}{4} \right)^{l_2}\nonumber\\
    &\quad \times\sqrt{\dfrac{(2l+1)!}{(2l_1)!(2l_2)!}}\sqrt{\dfrac{(2\lambda+1)!}{(2\lambda_1)!(2\lambda_2)!}}\, \nonumber\\
    &\quad \times  C_{l_1 \lambda_1 l'}^{0\,0\,0}C_{l_2 \lambda_2 \lambda'}^{0\,0\,0}
    \left \{ 
        \begin{array}{ccc}  l_1 &l_2 & l\\ \lambda_1& \lambda_2&\lambda \\ l' &\lambda' &  {L}
        \end{array}
    \right \}\, .\label{eqn:gaa_k=0}
\end{align}

The Wigner $9\mbox{-}j$ symbol embedded in $\tilde{g}_{\alpha' \alpha}^{0 l_1 l_2 \lambda_1 \lambda_2}$ has a set of triangle rules that indicate the following restrictions:
\begin{align}
    l' &\leqslant l_1+\lambda_1\, ,\label{eqn:l1Plusl2}\\
    \lambda' &\leqslant l_2+\lambda_2\, , \label{eqn:lmbd1Pluslmbd2}\\
    L &\leqslant l' + \lambda' \, .
\end{align}
The above inequalities and identities $l_1+l_2=l$ and $\lambda_1+\lambda_2=\lambda$ suggest 
\begin{equation}
  L \leqslant  l'+\lambda'\leqslant l+\lambda\, . 
  \label{eqn:RangeOflpPluslmbdp}
\end{equation}
On one hand, $l' + \lambda'$ of the permutation-transformed state can take any integer value between $L$ and $l+\lambda$, provided that $l' + \lambda'$ and $l+\lambda$ are both even or both odd as demanded by the conservation of parity. On the other hand, Eq.~\eqref{eqn:ScalTrans} requires $l' + \lambda'$ to take a fixed value: $l' + \lambda' = l + \lambda = \nu$.
This can only be achieved by limiting $L$ to two possible values: $\nu$ or $\nu-1$ (if $\nu \geqslant 2)$.

Forcing $l' + \lambda' = l + \lambda$ by limiting $L$ has another consequence: 
the equalities in inequalities \eqref{eqn:l1Plusl2} and \eqref{eqn:lmbd1Pluslmbd2} must hold. This is because $l'$ and $\lambda'$ are both non-negative and the upper bound of their sum is reached only when each of them reaches their own upper bound:
\begin{align}
    l'&=l_1+\lambda_1\, ,\\
    \lambda'&=l_2+\lambda_2\, .
\end{align}
Combined with Eq.~\eqref{eqn:pqaPHkEqs0}, the above equations ensure that Eq.~\eqref{eqn:Local_PH} satisfies:
\begin{align}
     &\braket{p'q' \alpha'|P_{123}|H }=
2 {p'}^{l'} {q'}^{\lambda'} \sum_{
    \substack{
    \alpha,\, l_1+l_2=l \\ 
    \lambda_1+\lambda_2=\lambda}} \psi_{\alpha} \,
  \tilde{g}_{\alpha' \alpha}^{0l_1 l_2 \lambda_1 \lambda_2}\, .
    \label{eqn:SmppqaPHkEqs0}
\end{align}

In practical calculations, one often applies ultraviolet regularizations. This implies 
that the high-momentum modes of the wave function $\braket{pq\alpha | H}$ are to be regulated.
The most convenient regulators in our approach are probably the ones invariant under any particle exchanges, e.g.,
\begin{equation}
    \braket{pq\alpha | H} =  \psi_\alpha p^l q^\lambda f_R \left(\dfrac{p^2+\frac{3}{4}q^2}{\Lambda^2} \right) \, ,
\end{equation}
where the regulator function $f_R(x)$ can be, for instance,
\begin{equation}
    f_R(x) = e^{-x^2}\, .
\end{equation}
With permutation-invariant regulators like this, the argumentation presented in this section can be straightforwardly repeated, leading to the same conclusion.

We summarize the criteria to identify a finite set of channels for expanding $\ket{H}$ with $m = n = 0$:
\begin{enumerate}
    \item For a given value of $\nu$, let $L$ be $\nu$ or $\nu - 1$ (if $\nu \geqslant 2$).
    \item Enumerate all combinations of $l\lambda$ that satisfy $l + \lambda = \nu$ and are allowed by the triangle rule imposed on $(l\lambda) L$.
    \item Enumerate all possible values of $stST$ and check $l + s + t = $ odd.
\end{enumerate}
Each set is denoted by $\Gamma_c$ and includes, by construction, one or more channels sharing identical values of $\nu LST$. They form finite-dimensional bases for expanding Eq.~\eqref{eqn:HPInv}.
But not all such bases yield permutation-invariant states; those that do are tabulated in Table.~\ref{tab:Subspace}. 

\begin{table*}
\centering
\renewcommand{\arraystretch}{1.0} 
\setlength{\tabcolsep}{15pt}
\caption{
The bases for expanding the permutation-invariant, zero-range $3N$ states. 
Each basis is denoted by $\Gamma_c$ where $c$ is a distinct integer and is distinguished by the values of 
$\nu L S T$.
The channels within a basis differ in $l\lambda s$. $t$ takes such a value that $l+s+t = \text{odd}$. See the text for explanations of the quantum numbers.
}
\begin{tabular}{c|ccc|cccc}    
    \hline
    \hline
    Label & ${L}$ & ${S}$ & ${T}$ & $l$ & $\lambda$ & $s$ \\
    \hline
    \multirow{2}{*}{$\Gamma_1$} & \multirow{2}{*}{0} & \multirow{2}{*}{1/2} & \multirow{2}{*}{1/2} & 0 & 0 & 0 \\
    & & & & 0 & 0 & 1 \\
    \hline
    \multirow{4}{*}{$\Gamma_2$} & \multirow{4}{*}{1} & \multirow{4}{*}{1/2} & \multirow{4}{*}{1/2} & 0 & 1 & 0 \\
    & & & & 0 & 1 & 1 \\
    & & & & 1 & 0 & 0 \\
    & & & & 1 & 0 & 1 \\
    \hline
    \multirow{2}{*}{$\Gamma_3$} & \multirow{2}{*}{1} & \multirow{2}{*}{1/2} & \multirow{2}{*}{3/2} & 0 & 1 & 0 \\
    & & & & 1 & 0 & 1 \\
    \hline
    \multirow{2}{*}{$\Gamma_4$} & \multirow{2}{*}{1} & \multirow{2}{*}{3/2} & \multirow{2}{*}{1/2} & 0 & 1 & 1 \\
    & & & & 1 & 0 & 1 \\
    \hline
    \multirow{2}{*}{$\Gamma_5$} & \multirow{2}{*}{1} & \multirow{2}{*}{1/2} & \multirow{2}{*}{1/2} & 1 & 1 & 0 \\
    & & & & 1 & 1 & 1 \\
    \hline
    \multirow{3}{*}{$\Gamma_6$} & \multirow{3}{*}{2} & \multirow{3}{*}{3/2} & \multirow{3}{*}{1/2} & 0 & 2 & 1 \\
    & & & & 1 & 1 & 1 \\
    & & & & 2 & 0 & 1 \\
    \hline
    \multirow{1}{*}{$\Gamma_7$} & \multirow{1}{*}{1} & \multirow{1}{*}{3/2} & \multirow{1}{*}{3/2} & 1 & 1 & 1 \\
    \hline
    \multirow{6}{*}{$\Gamma_8$} & \multirow{6}{*}{2} & \multirow{6}{*}{1/2} & \multirow{6}{*}{1/2} & 0 & 2 & 0 \\
    & & & & 0 & 2 & 1 \\
    & & & & 1 & 1 & 0 \\
    & & & & 1 & 1 & 1 \\
    & & & & 2 & 0 & 0 \\
    & & & & 2 & 0 & 1 \\
    \hline
    \multirow{3}{*}{$\Gamma_9$} & \multirow{3}{*}{2} & \multirow{3}{*}{1/2} & \multirow{3}{*}{3/2} & 0 & 2 & 0 \\
    & & & & 1 & 1 & 1 \\
    & & & & 2 & 0 & 0 \\
    \hline
    \hline
\end{tabular}\label{tab:Subspace}
\end{table*}

It will facilitate the discussions in next section to show in detail how Eq.~\eqref{eqn:HPInv} is solved.
Suppose that we use $\Gamma_c$ as the basis to expand $\ket{H}$:
\begin{equation}
    \braket{pq\alpha | H} = 
    \begin{cases}
        \psi_{\alpha}^{(c)} \, p^l q^\lambda \quad &\alpha \in \Gamma_c \, , \\
        0 \quad &\text{otherwise}\, .
    \end{cases}\label{eqn:dfnPsi_alpha}
\end{equation}
It then follows from Eq.~\eqref{eqn:SmppqaPHkEqs0} that the linear combinations constructed above span a subspace invariant under the action of $P_{123}$:
\begin{equation}
    \braket{pq\alpha | P_{123} | H} = 
    \begin{cases}
        p^{l} q^{\lambda} \displaystyle\sum_{\alpha' \in \Gamma_c} \mathscr{A}^{(c)}_{\alpha \alpha'} \psi_{\alpha'}^{(c)}
        \quad & \alpha \in \Gamma_c\, , \\
        0 \quad & \text{otherwise} \, .
    \end{cases}
\end{equation}
Here 
\begin{equation} 
\mathscr{A}^{(c)}_{\alpha \alpha'} \equiv 
2
\sum_{\substack{l_1+l_2=l' \\ 
    \lambda_1+\lambda_2=\lambda'}}
\tilde{g}_{\alpha \alpha'}^{0l_1l_2\lambda_1\lambda_2}\, .
\end{equation}

To ensure the permutation invariance of $\ket{H}$, the column vector $\psi_\alpha^{(c)}$ needs to be invariant under the transformation defined by $\mathscr{A}_{\alpha \alpha'}^{(c)}$:
\begin{align}    
\sum_{\alpha' \in \Gamma_c} \mathscr{A}^{(c)}_{\alpha  \alpha'}\psi_{\alpha'}^{(c)} = \psi_{\alpha}^{(c)} \quad \text{for}\; \alpha \in \Gamma_c\, .
\label{eqn:EigenEuqation}
\end{align}
In other words, $\mathscr{A}^{(c)}_{\alpha  \alpha'}$ needs to have an eigenvalue of $1$.
Equation~\eqref{eqn:RepP123} indicates that the nontrivial matrix elements of $\mathscr{A}_{\alpha \alpha'}^{(c)}$ are independent of $M_L$, $M_S$, or $M_T$:
\begin{equation}
    \mathscr{A}_{\alpha \alpha'}^{(c)} = 
    A_{l\lambda s\,,\, l'\lambda's'}^{(c)}
    \delta_{M_L M_L'} \delta_{M_S M_S'} \delta_{M_T M_T'}\, ,
    \label{eqn:DefRedcA}
\end{equation}
where each combination of $l\lambda s$ comes from a row on the right in Table~\ref{tab:Subspace} that belongs to $\Gamma_c$.
It follows that $\psi_\alpha^{(c)}$ depends only on $l\lambda s$:
\begin{equation}
    \psi_\alpha^{(c)} = \psi_{l\lambda s}^{(c)} \, .
\end{equation}
If the dimension of $\Gamma_c$ is $N_c$, as indicated by the number of rows of $\Gamma_c$ in Table~\ref{tab:Subspace}, the reduced matrix $A_{l\lambda s\,,\, l'\lambda's'}^{(c)}$ is a $N_c \times N_c$ matrix, and $\psi_{l\lambda s}^{(c)}$ is a vector with $N_c$ components. $A_{l\lambda s\,,\, l'\lambda's'}^{(c)}$ of each $\Gamma^{(c)}$ are shown in Appendix~\ref{app:MatAVecC} , together with the corresponding eigenvector $\psi^{(c)}$.

\subsection{\texorpdfstring{$m > 0$}{mgt0} or \texorpdfstring{$n > 0$}{ngt0}\label{sec:mnnonezero}}

When including the cases where the factors of $p^{2m}$ or $q^{2n}$ are presented in the wave function of $\ket{H}$, as defined in Eq.~\eqref{eqn:Def_H}, we are no longer restricted by Eqs.~\eqref{eqn:llmbdeqnu} and \eqref{eqn:Leqnu}.
However, the permutation-invariant subspaces can still be classified by the values of $\nu LST$ because they remain unchanged under $P_{123}$.
The invariant subspaces that have been accounted for--- those listed in Table~\ref{tab:Subspace}--- can now be excluded.
In addition, since we stop at the level of $\mathcal{O}(Q^2)$, $\nu = 2$ is all we need to work on.
With all these considerations, we tabulate the remaining partial waves in Table~\ref{tab:SubspaceII}.

\begin{table*}
\centering
\renewcommand{\arraystretch}{1.0} 
\setlength{\tabcolsep}{15pt}
\caption{
The remaining partial waves that will be used in the paper to construct $3N$ permutation-invariant states.
}
\begin{tabular}{c|ccc|cccc}    
    \hline
    \hline
    Label & ${L}$ & ${S}$ & ${T}$ & $l$ & $\lambda$ & $s$ \\
    \hline
    \multirow{1}{*}{$\Gamma_{10}$} & \multirow{1}{*}{0} & \multirow{1}{*}{3/2} & \multirow{1}{*}{1/2} & 0 & 0 & 1 \\
    \hline
    \multirow{1}{*}{$\Gamma_{11}$} & \multirow{1}{*}{0} & \multirow{1}{*}{1/2} & \multirow{1}{*}{3/2} & 0 & 0 & 0 \\
    \hline
    \multirow{2}{*}{$\Gamma_{13}$} & \multirow{2}{*}{0} & \multirow{2}{*}{1/2} & \multirow{2}{*}{1/2} & 1 & 1 & 0 \\
    & & & & 1 & 1 & 1 \\
    \hline
    \multirow{1}{*}{$\Gamma_{14}$} & \multirow{1}{*}{0} & \multirow{1}{*}{3/2} & \multirow{1}{*}{1/2} & 1 & 1 & 1 \\
    \hline
    \multirow{1}{*}{$\Gamma_{15}$} & \multirow{1}{*}{0} & \multirow{1}{*}{1/2} & \multirow{1}{*}{3/2} & 1 & 1 & 0 \\
    \hline
    \hline
\end{tabular}\label{tab:SubspaceII}
\end{table*}

For $\nu LST=\{2\, 0\, \frac{1}{2}\, \frac{1}{2}\}$, we find the following three possible components for $\ket{H}$:
\begin{equation}
    pq \ket{pq\Gamma_{13}}\, , \quad
    p^2\ket{pq\Gamma_1} \, ,\quad 
    q^2\ket{pq\Gamma_1} \, ,
\end{equation}
where $p^2\ket{pq \Gamma_1}$ or $q^2\ket{pq \Gamma_1}$ represents a different basis vector than $\ket{pq \Gamma_1}$ due to the factors of $p^2$ and $q^2$.
Because $\Gamma_1$ and $\Gamma_{13}$ are both two-dimensional, ignoring again $M_L$, $M_S$, or $M_T$, the dimension of this basis is six.
There are two other sets of $\nu LST$, each composed of two $\Gamma$'s.
For $\nu LST=\{2\, 0\, \frac{3}{2}\, \frac{1}{2}\}$: 
\begin{equation}
    pq \ket{pq\Gamma_{14}}\, , \quad p^2\ket{pq\Gamma_{10}}\, ,\quad q^2\ket{pq\Gamma_{10}}\, ,
\end{equation}
and for $\nu LST=\{2\, 0\, \frac{1}{2}\, \frac{3}{2}\}$:
\begin{equation}
    pq \ket{pq\Gamma_{15}}\, , \quad p^2\ket{pq\Gamma_{11}}\, ,\quad q^2\ket{pq\Gamma_{11}}\, .
\end{equation}

We proceed to expand Eq.~\eqref{eqn:HPInv} in each of these bases. 
For the sake of generality, we denote the two $\Gamma$'s in each basis as $\Gamma_x$ and $\Gamma_y$ and write $\ket{H}$ as
\begin{align}
    \braket{pq\alpha|H}=\begin{cases}
        \psi_{1\, l\lambda s}\, pq\, ,&\alpha\in\Gamma_{x}\, ,\\[8pt]
        \psi_{2\, l\lambda s}\,p^2\, ,& \alpha\in\Gamma_y\, ,\\[8pt]
        \psi_{3\, l\lambda s}\,q^2\, ,&\alpha\in\Gamma_y\, ,\\[8pt]
        0\, ,&\text{otherwise}.
    \end{cases}  \label{eqn:DefHGmxGmy}  
\end{align}
Applying $P_{123}$ to $\braket{pq\alpha|H}$  and collecting the corresponding coefficients of $p'q'$, $p'^2$, and $q'^2$, we obtain the following linear equation for $\ket{H}$:
\begin{align}
    \sum_{\alpha'\in\{\Gamma_{x},\Gamma_{y}\}}A_{i l\lambda s\,,\, jl'\lambda's'}^{(x,y)}\psi_{jl'\lambda's'}=\psi_{i l\lambda s}\, ,  \label{eqn:matAwp2q2}  
\end{align}
where $i, j = \{1, 2, 3\}$ and
\begin{align}
A_{i l\lambda s\,,\, jl'\lambda's'}^{(x,y)}=
\begin{pmatrix}
2(\tilde{g}_{\Gamma_{x}\Gamma_{x}}^{01001}+\tilde{g}_{\Gamma_x\Gamma_x}^{00110})&\dfrac{1}{2}\tilde{g}_{\Gamma_x'\Gamma_y}^{10000}&-\dfrac{2}{3}\tilde{g}_{\Gamma_x'\Gamma_y}^{10000}\\[8pt]
2\tilde{g}_{\Gamma_y'\Gamma_x}^{01010}&\dfrac{1}{2}\tilde{g}_{\Gamma_y'\Gamma_y}^{00000}&2\tilde{g}_{\Gamma_y'\Gamma_y}^{00000}\\[8pt]
2\tilde{g}_{\Gamma_y'\Gamma_x}^{00101}&\dfrac{9}{8}\tilde{g}_{\Gamma_y'\Gamma_y}^{00000}&\dfrac{1}{2}\tilde{g}_{\Gamma_y'\Gamma_y}^{00000}
    \end{pmatrix}\, .
\end{align}
We relegate again to Appendix~\ref{app:MatAVecC} the matrices $A$ and the permutation-invariant vectors $\ket{H}$ found in these bases.

\section{Three-nucleon contact interactions\label{sec:CntOpt}}

The $3N$ contact interactions are now constructed by forming the outer products of $\ket{H_c}$ and $\ket{H_d}$, as demonstrated in Eq.~\eqref{eqn:OutPrdtHiHj}. Priority is given to the states $\ket{H_c}$ and $\ket{H_d}$ constructed in Sec.~\ref{sec:mn0}. We explain the notation by assuming that this is the case: $\ket{H_c}$ and $\ket{H_d}$ are respectively from $\Gamma_c$ and $\Gamma_d$. It will become obvious how to extended the notation to include the invariant states constructed in Sec.~\ref{sec:mnnonezero}.

In order to conserve the total angular momentum $J$, isospin $T$, and parity $\mathcal{P}$, $\ket{H_c}$ and $\ket{H_d}$ must share identical values of these quantum numbers. Therefore, it is preferable to employ the coupled $LS$ scheme for the expansion of $\ket{H_c}$ and $\ket{H_d}$:
\begin{widetext}
    \begin{align}
       \ket{H_c} &= \int \dd q \dd p \, q^2 p^2 \sum_{\{l\lambda s\} \in \Gamma_c} \psi_{l\lambda s}^{(c)} \, p^l q^\lambda \ket{pq\; (l\lambda) {L_c} \left (s \frac{1}{2}\right )  {S_c}JM_J\left (t \frac{1}{2}\right ) T M_T}\, ,
    \end{align}
\end{widetext}

where $M_J$ is the $z$ component of the total angular momentum and the vector $\psi_{l\lambda s}^{(c)}$ can be found in Appendix~\ref{app:MatAVecC} for any given $\Gamma_c$.
The contact interactions
have the following generic structure:
\begin{equation}
    W^{JT\mathcal{P}}_{{S_c L_c}, {S_d L_d}} \propto \sum_{M_{J} M_T} \left(\ket{H_c} \bra{H_d} + h.c.\right)\, ,
\end{equation}
where $S_c L_c$ and $S_d L_d$ are the quantum numbers of $\ket{H_c}$ and $\ket{H_d}$, respectively, while $M_J$ and $M_T$ are shared by both states.
The summation over $M_J$ and $M_T$ guarantees that the resulting operator is both a scalar and an iso-scalar.

It is obvious that the nonvanishing matrix elements $W^{JT\mathcal{P}}_{{S_c L_c}, {S_d L_d}}$ are those between the channels that belong respectively to $\Gamma_c$ and $\Gamma_d$. 
For $\alpha' \in \Gamma_c$ and $\alpha \in \Gamma_d$ in the \emph{coupled} $LS$ scheme: 
\begin{align}
    &\braket{p'q'\alpha' | W^{JT\mathcal{P}}_{{S_c L_c}, {S_d L_d}} | pq\alpha} = \nonumber \\
    & \beta_{c, d}^J
    \psi^{(c)}_{l'\lambda' s'} \psi^{(d)}_{l\lambda s} 
    p'^{l'}q'^{\lambda'} p^l q^\lambda \delta_{J'J} \delta_{M_J' M_J}  \delta_{T' T} \delta_{M_T' M_T},
\end{align}
where $\beta_{c, d}^J$ is the coupling constant. Note that $J M_J T M_T$ ($J' M_J' T' M_T'$) are quantum numbers of $\alpha$ ($\alpha'$), not to be confused with those of $\ket{H_{c(d)}}$.
It is sufficient to show the reduced matrix elements of $W^{JT\mathcal{P}}_{{S_c L_c}, {S_d L_d}}$ as a $N_c \times N_d$ matrix:
\begin{equation}
\braket{p'q'\Gamma_c | W^{JT\mathcal{P}}_{{S_c L_c}, {S_d L_d}} | pq\Gamma_d} \equiv \beta_{c, d}^J \psi^{(c)}_{l'\lambda' s'} \psi^{(d)}_{l\lambda s} 
    p'^{l'}q'^{\lambda'} p^l q^\lambda. 
\end{equation}
Here $l'\lambda's'$ associated with the $i$th row of this matrix corresponds to the $i$th row of the basis $\Gamma_c$ in Table~\ref{tab:Subspace}, and $l\lambda s$ associated with the $j$th column corresponds to the $j$th row of $\Gamma_d$. 

The $P_{123}$-invariant states discussed in Sec.~\ref{sec:mnnonezero} have nontrivial factor of $p^2$ or $q^2$ in their momentum-space wave functions and
are listed in the second part of Appendix~\ref{app:MatAVecC}. 
They all have components from two $\Gamma$'s instead of one. 
We denote by $U$ the $3N$ potentials that involve these permutation-invariant states.
For instance, the product of the state defined by Eq.~\eqref{eqn:psiprime131} and $\ket{pq \Gamma_1}$ is written as
\begin{align}
    &\begin{pmatrix}
        \bra{p'q'\Gamma_1}\\
        \bra{p'q'\Gamma_{13}}
    \end{pmatrix} U^{\frac{1}{2}\frac{1}{2}+}_{{{}^2S},\,{{}^2S}}\ket{pq\Gamma_1} \, ,
\end{align}
which is a matrix of the dimensions: $(N_{\Gamma_1} + N_{\Gamma_{13}}) \times N_{\Gamma_1} = 4 \times 2$.

We present in the following the $3N$ contact potentials up to  $\mathcal{O}(Q^2)$. 
They are split into two groups, each of which has a definite isospin: $T = 1/2$ and $3/2$.
Because we stop at the level of $\mathcal{O}(Q^2)$, not all of the invariant $3N$ states listed in Appendix~\ref{app:MatAVecC} will be utilized here.

\subsection{\texorpdfstring{$T = \frac{1}{2}$}{T12}}

\noindent $\{\boldsymbol{J\,T\,\mathcal{P}\}=\{\frac{1}{2}\,\frac{1}{2}\,+\}}$: \par

In our notation, the well-known $\mathcal{O}(Q^0)$ $3N$ contact potential is expressed as follows:
\begin{align}
    \bra{p'q'\Gamma_1}W^{\frac{1}{2}\frac{1}{2}+}_{{{}^2S},\,{{}^2S}}\ket{pq\Gamma_1}& = \beta_{1, 1}^{\frac{1}{2}} \begin{pmatrix}
        \phantom{-}1 & -1\\
        -1&\phantom{-}1
\end{pmatrix}\, .
\label{eqn:3NFMat2S12}
\end{align}

Next, we list all the  $\mathcal{O}(Q^2)$ $3N$ contact potentials:
\begin{align}
    &\bra{p'q'\Gamma_1} U^{\frac{1}{2}\frac{1}{2}+}_{{{}^2S},\,{{}^2S}}\ket{pq\Gamma_1}\nonumber\\
    &=\gamma_{1,1}^{\frac{1}{2}}\begin{pmatrix}
        \phantom{-}1&-1\\
        -1&\phantom{-}1
    \end{pmatrix}(p'^2+\frac{3}{4}q'^2+p^2+\dfrac{3}{4}q^2)\, .
\end{align}

\begin{align}
    &\begin{pmatrix}
        \bra{p'q'\Gamma_1}\\
        \bra{p'q'\Gamma_{13}}
    \end{pmatrix} U^{\frac{1}{2}\frac{1}{2}+}_{{{}^2S},\,{{}^2S}}\ket{pq\Gamma_1}\hspace{3.5cm}\nonumber\\
    &\,\,\,=\gamma_{1,13}^{\frac{1}{2}}\begin{pmatrix}
        -p'^2+\dfrac{3}{4}q'^2&p'^2-\dfrac{3}{4}q'^2\\[8pt]
        -p'^2+\dfrac{3}{4}q'^2&p'^2-\dfrac{3}{4}q'^2\\[8pt]
        \phantom{-}p'q'&-p'q'\\[8pt]
        -p'q'&\phantom{-}p'q'
    \end{pmatrix}\, .
\end{align}

\begin{align}
    \bra{p'q'\Gamma_1}W^{\frac{1}{2}\frac{1}{2}+}_{^2S,\, ^2P}\ket{pq\Gamma_5}& = \beta_{1, 5}^{\frac{1}{2}}  \begin{pmatrix}
        -pq & -pq\\
        \phantom{-}pq & \phantom{-}pq\\
    \end{pmatrix}\, .
\end{align}
\begin{align}
    \bra{p'q'\Gamma_1}W^{\frac{1}{2}\frac{1}{2}+}_{^2S,\, ^4D}\ket{pq\Gamma_6}&=  \beta_{1,6}^{\frac{1}{2}}\begin{pmatrix}
        \phantom{-}\frac{3}{4}q^2 & -\frac{\sqrt{10}}{2}pq & -p^2\\[0.5ex]
        -\frac{3}{4}q^2 &\phantom{-} \frac{\sqrt{10}}{2}pq & \phantom{-}p^2\\
    \end{pmatrix}  \, .
\end{align}

\noindent 
$\boldsymbol{\{J\,T\,\mathcal{P}\}=\{\frac{1}{2}\,\frac{1}{2}\,-\}}$:
\par
\begin{align}
    \bra{p'q'\Gamma_4}W^{\frac{1}{2}\frac{1}{2}-}_{^4P,\, ^4P}\ket{pq\Gamma_4} &= \beta_{4,4}^{\frac{1}{2}}\begin{pmatrix}
        \frac{3}{4}q'q & \frac{\sqrt{3}}{2}q'p\\[0.5ex]
        \frac{\sqrt{3}}{2}p'q&p'p
    \end{pmatrix}  \, .
\end{align}
\begin{align}
&\bra{p'q'\Gamma_2}W^{\frac{1}{2}\frac{1}{2}-}_{^2P,\, ^2P}\ket{pq\Gamma_2} \nonumber \\
&= \beta_{2,2}^{\frac{1}{2}}\begin{pmatrix}
        \phantom{-}\frac{3}{4}q'q & \phantom{-}\frac{3}{4}q'q& \phantom{-}\frac{\sqrt{3}}{2}q'p & -\frac{\sqrt{3}}{2}q'p\\[0.5ex]
        \phantom{-}\frac{3}{4}q'q &\phantom{-} \frac{3}{4}q'q& \phantom{-}\frac{\sqrt{3}}{2}q'p & -\frac{\sqrt{3}}{2}q'p\\[0.5ex]
        \phantom{-}\frac{ \sqrt{3}}{2}p'q & \phantom{-}\frac{\sqrt{3}}{2}p'q& \phantom{-}p'p & -p'p\\[0.5ex]
        -\frac{\sqrt{3}}{2}p'q & -\frac{\sqrt{3}}{2}p'q& -p'p & \phantom{-}p'p\\
    \end{pmatrix} \, .
\end{align}
\begin{align}
&\bra{p'q'\Gamma_4}W^{\frac{1}{2}\frac{1}{2}-}_{^4P,\, ^2P}\ket{pq\Gamma_2} \nonumber\\
    &= \beta_{4,2}^{\frac{1}{2}}\begin{pmatrix}
        -\frac{3}{4}q'q & -\frac{3}{4}q'q& -\frac{\sqrt{3}}{2}q'p & \frac{\sqrt{3}}{2}q'p\\[0.5ex]
        -\frac{\sqrt{3}}{2}p'q& -\frac{\sqrt{3}}{2}p'q & -p'p &p'p\\
        \end{pmatrix} \, .
\end{align}

\noindent
$\boldsymbol{\{J\,T\,\mathcal{P}\}=\{\frac{3}{2}\,\frac{1}{2}\,-\}}$: \par
\begin{align}
&\bra{p'q'\Gamma_4}W^{\frac{3}{2}\frac{1}{2}-}_{^4P, \, ^4P}\ket{pq\Gamma_4} 
= \beta_{4,4}^{\frac{3}{2}}\begin{pmatrix}
        \frac{3}{4}q'q & \frac{\sqrt{3}}{2}q'p\\[0.5ex]
        \frac{\sqrt{3}}{2}p'q&p'p
    \end{pmatrix}  \, .
\end{align}
\begin{align}
&\bra{p'q'\Gamma_2}W^{\frac{3}{2}\frac{1}{2}-}_{^2P, \, ^2P}\ket{pq\Gamma_2} \nonumber \\
&= \beta_{2,2}^{\frac{3}{2}}\begin{pmatrix}
        \phantom{-}\frac{3}{4}q'q & \phantom{-}\frac{3}{4}q'q& \phantom{-}\frac{\sqrt{3}}{2}q'p & -\frac{\sqrt{3}}{2}q'p\\[0.5ex]
        \phantom{-}\frac{3}{4}q'q &\phantom{-} \frac{3}{4}q'q& \phantom{-}\frac{\sqrt{3}}{2}q'p & -\frac{\sqrt{3}}{2}q'p\\[0.5ex]
        \phantom{-}\frac{ \sqrt{3}}{2}p'q & \phantom{-}\frac{\sqrt{3}}{2}p'q& \phantom{-}p'p & -p'p\\[0.5ex]
        -\frac{\sqrt{3}}{2}p'q & -\frac{\sqrt{3}}{2}p'q& -p'p & \phantom{-}p'p\\
    \end{pmatrix} \, .
\end{align}
\begin{align}
&\bra{p'q'\Gamma_4}W^{\frac{3}{2}\frac{1}{2}-}_{^4P, \, ^2P}\ket{pq\Gamma_2} \nonumber \\
&= \beta_{4,2}^{\frac{3}{2}}\begin{pmatrix}
        -\frac{3}{4}q'q & -\frac{3}{4}q'q& -\frac{\sqrt{3}}{2}q'p & \frac{\sqrt{3}}{2}q'p\\[0.5ex]
        -\frac{\sqrt{3}}{2}p'q& -\frac{\sqrt{3}}{2}p'q & -p'p &p'p\end{pmatrix} \, .
\end{align}

\noindent
$\boldsymbol{\{J\,T\,\mathcal{P}\}=\{\frac{5}{2}\,\frac{1}{2}\,-\}}$: \par
\begin{align}
    \bra{p'q'\Gamma_4}W^{\frac{5}{2}\frac{1}{2}-}_{^4P, \, ^4P}\ket{pq\Gamma_4} &= \beta_{4,4}^{\frac{5}{2}}\begin{pmatrix}
        \frac{3}{4}q'q & \frac{\sqrt{3}}{2}q'p\\[0.5ex]
        \frac{\sqrt{3}}{2}p'q&p'p
    \end{pmatrix}\, .
\end{align}

\subsection{\texorpdfstring{$T = \frac{3}{2}$}{T32}}

\noindent
$\boldsymbol{\{J\,T\,\mathcal{P}\}=\{\frac{1}{2}\,\frac{3}{2}\,-\}}$: \par
\begin{align}
&\bra{p'q'\Gamma_3}{W}^{\frac{1}{2}\frac{3}{2}-}_{^2P,\,^2P}\ket{pq\Gamma_3} 
= \beta_{3,3}^{\frac{1}{2}}
\begin{pmatrix}
    \frac{3}{4}q'q & \frac{\sqrt{3}}{2}q'p\\[0.5ex]
    \frac{\sqrt{3}}{2}p'q&p'p
\end{pmatrix} \, .
\end{align}

\noindent
$\boldsymbol{\{J\,T\,\mathcal{P}\}=\{\frac{3}{2}\,\frac{3}{2}\,-\}}$: \par
\begin{equation}
    \bra{p'q'\Gamma_3}{W}^{\frac{3}{2}\frac{3}{2}-}_{^2P, \, ^2P}\ket{pq\Gamma_3} = \beta_{3,3}^{\frac{3}{2}}\begin{pmatrix}
        \frac{3}{4}q'q & \frac{\sqrt{3}}{2}q'p\\[0.5ex]
        \frac{\sqrt{3}}{2}p'q&p'p
    \end{pmatrix} \, .
\end{equation}

\section{Plane-wave basis\label{sec:plane}}

In order to compare with the potentials derived from other approaches, e.g., directly from the chiral Lagrangian~\cite{Girlanda:2011fh}, it is necessary to discuss how to transform $3N$ contact potentials from the plane-wave basis, i.e., three-momentum eigen states to the partial-wave basis used in this paper.
We review in this section the general framework of this transformation.

The $3N$ potentials expressed in the plane-wave basis are often decomposed into three components that can be changed to one another by $P_{123}$:
\begin{equation}
V_{\text{3CT}} = V^{(1)}_{\text{3CT}} 
+ V^{(2)}_{\text{3CT}}
+ V^{(3)}_{\text{3CT}} \, .
\end{equation}
where
\begin{align}
V^{(1)}_{\text{3CT}} &= P_{123}V^{(3)}_{\text{3CT}}P_{123}^{-1}\, , \\
V^{(2)}_{\text{3CT}}
&=P_{123}^{-1}V^{(3)}_{\text{3CT}}P_{123}\, .
\end{align}
Given the expression of $V_{\text{3CT}}^{(3)}$ in the plane-wave basis
\begin{equation*}
    V_{\text{3CT}}^{(3)}(\vec{p'},\vec{q'};\vec{p},\vec{q}) \, ,
\end{equation*}
where $\vec{p}$ and $\vec{q}$ ($\vec{p}'$ and $\vec{q}'$) are the incoming (outgoing) Jacobi three-momenta and $V_{\text{3CT}}^{(3)}$ is in general a matrix in spin and isospin space,
we wish to find its matrix elements in the partial-wave basis.
If our contact potentials listed in Sec.~\ref{sec:CntOpt} are complete up to $\mathcal{O}(Q^2)$, any $V_{\text{3CT}}^{(3)}$ within $\mathcal{O}(Q^2)$ can be decomposed into a linear combination of those potentials.

Because the operators developed in this paper are by construction built on antisymmetrized states, the comparison is meaningful only if $V_{\text{3CT}}$ is projected onto an antisymmetrized basis.
Specifically, we consider the following matrix element:
    \begin{align}
    &\braket{p'q'\alpha'|\mathcal{A}_{123}V_{\text{3CT}}\mathcal{A}_{123}|pq\alpha} \nonumber \\
    =&\bra{p'q'\alpha'}(1+P_{123}+P_{123}^{-1})\dfrac{V_{\text{3CT}}}{9} (1+P_{123}+P_{123}^{-1})\ket{pq\alpha}\nonumber \\
    =&\bra{p'q'\alpha'}(1+P_{123}+P_{123}^{-1})\dfrac{V^{(3)}_{\text{3CT}}}{3} (1+P_{123}+P_{123}^{-1})\ket{pq\alpha}
    \, ,\label{eqn:Anti3NF}
    \end{align}
where the three-body antisymmetrizer~\cite{Hebeler:2020ocj} is defined by 
\begin{equation}
    \mathcal{A}_{123}=\frac{1-P_{12}}{2}\dfrac{1+P_{123}+P_{123}^{-1}}{3}\, ,
\end{equation}
and we have used the fact that the Jacobi basis is antisymmetrized between nucleon 1 and 2.

We need to deal with the following integral
\begin{widetext}
    \begin{align}
     \braket{p'q'\alpha'|(1+P_{123}+P_{123}^{-1})V^{(3)}_{\text{3CT}}(1+P_{123}+P_{123}^{-1})|pq\alpha}\, \hspace{4cm}\nonumber\\    
    =\int \dfrac{\dd^3\vec{a}'}{(2\pi)^3} \dfrac{\dd^3\vec{b}'}{(2\pi)^3} \dfrac{\dd^3\vec{a}}{(2\pi)^3} \dfrac{\dd^3\vec{b}}{(2\pi)^3}
    V^{(3)}_{\text{3CT}}(\vec{a}',\vec{b}';\vec{a},\vec{b})\bra{p'q'\alpha'}1+P_{123}+P_{123}^{-1}\ket{\vec{a}'\vec{b}'} \bra{\vec{a}\vec{b}}1+P_{123}+P_{123}^{-1}\ket{pq\alpha}\, .
    \label{eqn:Anti3NFToJacobi} 
    \end{align}
\end{widetext}

Because the three-momenta are separated from the spin and isospin degrees of freedom in the plane-wave basis, $\braket{\vec{a}\vec{b}|P_{123}|pq\alpha}$ is more easily understood in the uncoupled $LS$ partial-wave basis.
Although the action of $P_{123}$ in spin and isospin space can be read off from Eqs.~\eqref{eqn:RepP123}, \eqref{eqn:Gpqx}, and \eqref{eqn:Smallg}, it is reproduced here for clarity:
\begin{align}
    &\braket{\left (s'\frac{1}{2}\right )S'M_S'|P_{123}|\left(s\frac{1}{2}\right)SM_S}\nonumber\\
    &=\braket{\left (s'\frac{1}{2}\right )S'M_S'|P_{123}^{-1}|\left(s\frac{1}{2}\right)SM_S}\nonumber\\
    &=(-1)^{s}\sqrt{\hat{s}\hat{s}'}\delta_{SS'}\delta_{M_SM_S'}\begin{Bmatrix}
        \frac{1}{2}&\frac{1}{2}&s'\\[0.5ex]
        \frac{1}{2}&S&s
    \end{Bmatrix}\, \label{eqn:PMatInSpin}.
\end{align}
The isospin expression is entirely analogous to the spin one and is therefore omitted. The action of $P_{123}$ on the orbital motion is more nontrivial:
\begin{align}
    &\braket{pq(l\lambda)LM_L|P_{123}|\vec{ab}}=\braket{pq(l\lambda)LM_L|P_{123}^{-1}|\vec{ab}}\nonumber\\
    &=(2\pi)^3\sum_{l_{ab}\lambda_{ab}}\braket{pq(l\lambda)LM_L|P_{123}|ab(l_{ab}\lambda_{ab})LM_L}\nonumber\\
    &\qquad \quad \times \mathcal{Y}_{l_{ab}\lambda_{ab}}^{*LM_L}(\hat{\vec{a}},\hat{\vec{b}})\, ,\label{eqn:llmbdP123ab}
\end{align}
where $\mathcal{Y}_{l\lambda}^{LM_L}$ is the coupled spherical harmonic~\cite{Hebeler:2020ocj}:
\begin{equation}
  \mathcal{Y}_{l\lambda}^{LM_L}(\hat{\vec{a}},\hat{\vec{b}})=\sum_{m_lm_{\lambda}}C_{l\lambda L}^{m_lm_{\lambda}M_L}Y_{lm_l}(\hat{\vec{a}})Y_{\lambda m_{\lambda}}(\hat{\vec{b}})\, .
\end{equation}
Because the spin and isospin parts have been clarified,  $\braket{pq(l\lambda)LM_L|P_{123}|ab(l_{ab}\lambda_{ab})LM_L}$ can be easily identified from the orbital expression in Eqs.~\eqref{eqn:RepP123}, \eqref{eqn:Gpqx}, and \eqref{eqn:Smallg}.

The example we use to demonstrate the transformation is the $\beta_1$ operator defined in Eq.~\eqref{eqn:Leading3NF}:
\begin{equation}
    V_{\beta_1}^{(3)}(\vec{p'},\vec{q'};\vec{p},\vec{q})=\beta_1 \, .
\end{equation}
Evaluating the integral in Eq.~\eqref{eqn:Anti3NFToJacobi} is often simplified by noting that $V_\text{3CT}$ is a polynomial in momenta, therefore the angular integration over $\vec{a}'$ and $\vec{b}'$ can be separated from that over $\vec{a}$ and $\vec{b}$ and each of them is easily performed.
For $V_{\beta_1}^{(3)}$, the angular integration over $\vec{a}$ and $\vec{b}$ yields
\begin{equation}
    \int \dd \Omega_{\hat{\vec{a}}}\dd\Omega_{\hat{\vec{b}}}\mathcal{Y}_{l_{ab}\lambda_{ab}}^{LM_L}(\hat{\vec{a}},\hat{\vec{b}})=4\pi\delta_{l_{ab}0}\delta_{\lambda_{ab}0}\, ,
\end{equation}
which vanishes except for $l_{ab}=\lambda_{ab}=L=0$.
Using Eqs.~\eqref{eqn:RepP123}, \eqref{eqn:Gpqx}, and \eqref{eqn:Smallg}, 
we can further deduce that $l$, $\lambda$, $l'$, and $\lambda'$ in Eq.~\eqref{eqn:Anti3NFToJacobi}  must all be zero.

In addition, one can deduce $S=T=\frac{1}{2}$ either by direct computation of the likes of Eq.~\eqref{eqn:PMatInSpin} or by the antisymmetry requirement of $\mathcal{A}_{123}\ket{p'q'\alpha'}$ and $\mathcal{A}_{123}\ket{pq\alpha}$.

In conclusion, the only nonvanishing matrix elements of $V_{\beta_1}$ are between the partial waves of $\Gamma_1$ as defined in Table~\ref{tab:Subspace}. 
The final result is expressed as 
\begin{align}
\braket{p'q'\Gamma_1|\mathcal{A}_{123}V_{\beta_1}\mathcal{A}_{123}|pq\Gamma_1}=
\beta_1\dfrac{3}{8\pi^4}\begin{pmatrix}
        \phantom{-}1&-1\\[0.5ex]
        -1&\phantom{-}1
    \end{pmatrix}\, .
\end{align}
This is clearly the same operator as $W^{\frac{1}{2}\frac{1}{2}+}_{{{}^2S},\,{{}^2S}}$ defined in Eq.~\eqref{eqn:3NFMat2S12}, up to a constant.
Other momentum-independent operators, $V_{\beta_{2-6}}$ in Eq.~\eqref{eqn:Leading3NF} can be found to be proportional to $W^{\frac{1}{2}\frac{1}{2}+}_{{{}^2S},\,{{}^2S}}$ by similar computations.

\section{Summary and outlook}\label{sec:conclusion}

We have shown a machinery for building $3N$ contact interactions that can be used in chiral EFT and Pionless EFT.
Its advantage is that the redundancy among the $3N$ operators originating from exchanges of the nucleons as identical fermions is eliminated by construction.
The center of this machinery is a set of permutation-invariant, zero-range $3N$ states expanded in the Jacobi partial-wave basis. 
The $3N$ contact potentials will then be assembled as the products of these zero-range states.

The partial-wave basis simplifies the task because the cyclic permutation operator $P_{123}$ is block-diagonal in it.
The interplay of permutation invariance and zero-range nature enabled us to identify an array of finite-dimensional subspaces for $P_{123}$.
Once the subspaces are tabulated, solving the eigenvalues of $P_{123}$ is straightforward. 
If an eigenvalue of $1$ is found, its eigenvector will be one of the permutation-invariant states sought after. 
We have compiled all the zero-range states up to $\mathcal{O}(Q^2)$ in the Appendix.
The Jacobi partial-wave representation of $3N$ forces can be useful in many applications, including the Faddeev-Yakubovsky equations in the partial-wave basis, no-core shell model calculations~\cite{Nogga:2000vq, Barrett:2004qsq, Konig:2016utl, Konig:2019xxk, Navratil:2009ut, Stetcu:2006ey,Navratil:2007we}.

In the resulting $3N$ contact potentials, we have found $13$ independent $\mathcal{O}(Q^2)$ operators, identical to the number of independent operator reported in Ref.~\cite{Girlanda:2011fh}. 
To facilitate a more detailed comparison between these two sets of operators, we discussed how a $3N$ contact potential is transformed from the plane-wave basis to the momentum-space Jacobi basis, using the spin isospin independent potential $V_{\beta_1}$ as the example.
However the task of completely mapping our operators to those of Ref.~\cite{Girlanda:2011fh} is quite involved, and falls outside the scope of the present paper. 

The machinery presented in the paper focuses on removing redundancy due to the exchange of identical fermions.
It should be considered complementary to the standard Lagrangian approach rather than a self-contained alternative.
Therefore, it is not surprising that there are limitations to our method.
The transformation between the plane-wave basis and the partial-wave basis discussed in Sec.~\ref{sec:plane}  provides a foundation for the two methods to complement each other.
For instance, Lorentz invariance may impose a correlation among the contact $3N$ forces found in the paper.
Because Lorentz transformation is often implemented in the plane-wave basis, one can learn how such correlations are mapped in the partial-wave basis using this basis transformation.
The equation of motion, as mentioned earlier, may also further reduce the number of independent operators.
But this appears to be a less serious concern for operators at $\mathcal{O}(Q^2)$ in light of the experience of building contact $NN$ potentials.
A previous work on Pionless EFT~\cite{Beane:2000fi} suggests that without any energy-dependent potentials, the momentum-dependent operators will not develop equation-of-motion redundancy until $\mathcal{O}(Q^4)$.

\acknowledgments

We thank Luca Girlanda for useful discussions.
This work was supported by the National Natural Science Foundation of China (NSFC) under the Grants
Nos. 12275185 and 12335002.

\appendix

\section{Permutation-invariant \texorpdfstring{$3N$}{3N} states\label{app:MatAVecC}}
We show here some of the details related to the permutation-invariant zero-range $3N$ states constructed in Sec.~\ref{sec:3NZeroRangeState}. Starting with those states with $m = n = 0$, considered in Sec.~\ref{sec:mn0}, we list all the reduced matrices $A^{(c)}_{l\lambda s,l\lambda's'}$ defined in Eq.~\eqref{eqn:DefRedcA} and the invariant vector $\psi^{(c)}=\psi^{(c)}_{l\lambda s}$ under each $\Gamma_c$ of the Table.~\ref{tab:Subspace}. The columns and rows of $\mathbb{A}^{(c)}$, as well as the columns of $\psi^{(c)}$, are ordered according to the sequence of $\{l\lambda s\}$ listed in $\Gamma_c$. More specifically, $l'\lambda's'$  corresponds to the $j$th row of $\Gamma_c$ and is associated with the $j$th column of $\mathbb{A}^{(c)}$;
and $l\lambda s$ corresponds to the $i$th row of $\Gamma_c$ and is associated with the $i$th row of $\mathbb{A}^{(c)}$:
\begin{equation}
\mathbb{A}^{(c)}_{ij} = A^{(c)}_{l\lambda s,l'\lambda's'} \, .  
\end{equation}

\begin{widetext}
\begin{enumerate}
    \item $\Gamma_1: \{\nu L S T\}=\{0\, 0\, \frac{1}{2}\,\frac{1}{2}\}$
\begin{align}
\renewcommand{\arraystretch}{2.5} 
\mathbb{A}^{(1)} = 
\begin{pmatrix}
\phantom{-}\dfrac{1}{4} & -\dfrac{3}{4} \\
-\dfrac{3}{4} & \phantom{-}\dfrac{1}{4} 
\end{pmatrix}\, ,
\end{align}
and 
\begin{align}
\psi^{(1)} = 
\begin{pmatrix}
    -1 & 1
\end{pmatrix}^T \, .
\end{align}

 \item $\Gamma_2: \{\nu L S T\}=\{1\, 1\, \frac{1}{2}\, \frac{1}{2}\}$
\begin{equation}
\renewcommand{\arraystretch}{2.5} 
    \mathbb{A}^{(2)} =
    \begin{pmatrix}
        -\dfrac{1}{8} & \phantom{-}\dfrac{3}{8} & \phantom{-}\dfrac{3\sqrt{3}}{16} &-\dfrac{3\sqrt{3}}{16}\\
        
        \phantom{-}\dfrac{3}{8} & -\dfrac{1}{8} & \phantom{-}\dfrac{3\sqrt{3}}{16} &-\dfrac{3\sqrt{3}}{16} \\
        
        \phantom{-}\dfrac{\sqrt{3}}{4} & \phantom{-}\dfrac{\sqrt{3}}{4} & -\dfrac{1}{8} &-\dfrac{3}{8}\\
        
        -\dfrac{\sqrt{3}}{4} & -\dfrac{\sqrt{3}}{4} & -\dfrac{3}{8} &-\dfrac{1}{8} \\
    \end{pmatrix}
\end{equation}
and 
\begin{equation}
    \psi^{(2)} =
    \begin{pmatrix}
        -\dfrac{\sqrt{3}}{2} & -\dfrac{\sqrt{3}}{2}& -1&1
    \end{pmatrix}^T \, .
\end{equation}
    \item $\Gamma_3: \{\nu L S T\}=\{1\, 1\, \frac{1}{2}\, \frac{3}{2}\}$
\begin{equation}
\renewcommand{\arraystretch}{2.5} 
\mathbb{A}^{(3)} = 
\begin{pmatrix}
    \dfrac{1}{4} & \dfrac{3\sqrt{3}}{8} \\
    \dfrac{\sqrt{3}}{2} & \dfrac{1}{4} 
\end{pmatrix}
\end{equation}
and 
\begin{equation}
\psi^{(3)} = 
\begin{pmatrix}
    \dfrac{\sqrt{3}}{2} & 1
\end{pmatrix}^T\, .
\end{equation}

    \item $\Gamma_4: \{\nu L S T\}=\{1\, 1\, \frac{3}{2}\, \frac{1}{2}\}$
\begin{equation}
\renewcommand{\arraystretch}{2.5} 
\mathbb{A}^{(4)}=\begin{pmatrix}
    \dfrac{1}{4} & \dfrac{3\sqrt{3}}{8} \\
    \dfrac{\sqrt{3}}{2} & \dfrac{1}{4} 
\end{pmatrix}
\end{equation}
and 
\begin{equation}
\psi^{(4)}=\begin{pmatrix}
    \dfrac{\sqrt{3}}{2} & 1
\end{pmatrix}^T  \, .
\end{equation}

    \item $\Gamma_5: \{\nu L S T\}=\{2\, 1\, \frac{1}{2}\, \frac{1}{2}\}$
\begin{equation}
   \renewcommand{\arraystretch}{2.5} 
    \setlength{\arraycolsep}{8pt} 
    \mathbb{A}^{(5)}=\begin{pmatrix}
        \dfrac{1}{4} & \dfrac{3}{4} \\
        \dfrac{3}{4} & \dfrac{1}{4}
    \end{pmatrix}\, , 
\end{equation}
and 
\begin{equation}
    \psi^{(5)}=\begin{pmatrix}
        1 & 1
    \end{pmatrix}^T \, .
\end{equation}

    \item $\Gamma_6: \{\nu L S T\}=\{2\, 2\, \frac{3}{2}\, \frac{1}{2}\}$
\begin{equation}
   \renewcommand{\arraystretch}{2.5} 
    \mathbb{A}^{(6)}=\begin{pmatrix}
        -\dfrac{1}{8} & -\dfrac{9\sqrt{10}}{80} & -\dfrac{9}{32}  \\
        -\dfrac{\sqrt{10}}{4} & \phantom{-} \dfrac{1}{4} & \phantom{-} \dfrac{3\sqrt{10}}{16} \\
        -\dfrac{1}{2} & \phantom{-}\dfrac{3\sqrt{10}}{20} & -\dfrac{1}{8} 
    \end{pmatrix} 
\end{equation}
and 
\begin{equation}
    \psi^{(6)}=\begin{pmatrix}
        -\dfrac{3}{4} & \dfrac{\sqrt{10}}{2}& 1
    \end{pmatrix}^T \, .
\end{equation}

        \item $\Gamma_7: \{\nu L S T\}=\{2\, 1\, \frac{3}{2}\, \frac{3}{2}\}$
\begin{equation}
\renewcommand{\arraystretch}{2.5} 
\mathbb{A}^{(7)}= \left (1\right )\, .
\end{equation}
and 
\begin{equation}
\renewcommand{\arraystretch}{2.5} 
\psi^{(7)}= \left (1\right )\, .
\end{equation}

    \item $\Gamma_8: \{\nu L S T\}=\{2\, 2\, \frac{1}{2}\, \frac{1}{2}\}$
\begin{equation}
\renewcommand{\arraystretch}{2.5} 
\mathbb{A}^{(8)}=\begin{pmatrix}
\phantom{-}\dfrac{1}{16} & -\dfrac{3}{16} & -\dfrac{9\sqrt{10}}{160} & \phantom{-}\dfrac{9\sqrt{10}}{160} & \phantom{-}\dfrac{9}{64} & -\dfrac{27}{64} \\
-\dfrac{3}{16} & \phantom{-}\dfrac{1}{16} & -\dfrac{9\sqrt{10}}{160} & \phantom{-}\dfrac{9\sqrt{10}}{160} & -\dfrac{27}{64} & \phantom{-}\dfrac{9}{64} \\
-\dfrac{\sqrt{10}}{8} & -\dfrac{\sqrt{10}}{8} & -\dfrac{1}{8} & -\dfrac{3}{8} & \phantom{-}\dfrac{3\sqrt{10}}{32} & \phantom{-}\dfrac{3\sqrt{10}}{32} \\
\phantom{-}\dfrac{\sqrt{10}}{8} & \phantom{-}\dfrac{\sqrt{10}}{8} & -\dfrac{3}{8} & -\dfrac{1}{8} & -\dfrac{3\sqrt{10}}{32} & -\dfrac{3\sqrt{10}}{32} \\
\phantom{-}\dfrac{1}{4} & -\dfrac{3}{4} & \phantom{-}\dfrac{3\sqrt{10}}{40} & -\dfrac{3\sqrt{10}}{40} & \phantom{-}\dfrac{1}{16} & -\dfrac{3}{16} \\
-\dfrac{3}{4} & \phantom{-}\dfrac{1}{4} & \phantom{-}\dfrac{3\sqrt{10}}{40} & -\dfrac{3\sqrt{10}}{40} & -\dfrac{3}{16} & \phantom{-}\dfrac{1}{16}
\end{pmatrix}
\end{equation}
and 
\begin{align}
    \psi^{(8)}&=\begin{pmatrix}
        0&-\dfrac{3}{4} &\dfrac{\sqrt{10}}{4}&-\dfrac{\sqrt{10}}{4}& 1 &0
    \end{pmatrix}^T \, ,\\
    \psi'^{(8)}&=\begin{pmatrix}
        -\dfrac{3}{4}&0 &\dfrac{\sqrt{10}}{4}&-\dfrac{\sqrt{10}}{4} &0 &1
    \end{pmatrix}^T \, ,
\end{align}

        \item $\Gamma_9: \{\nu L S T\}=\{2\, 2\, \frac{1}{2}\, \frac{3}{2}\}$
\begin{equation}
   \renewcommand{\arraystretch}{2.5} 
    \mathbb{A}^{(9)}=\begin{pmatrix}
        -\dfrac{1}{8} & -\dfrac{9\sqrt{10}}{80} & -\dfrac{9}{32}  \\
        -\dfrac{\sqrt{10}}{4} & \phantom{-} \dfrac{1}{4} & \phantom{-} \dfrac{3\sqrt{10}}{16} \\
        -\dfrac{1}{2} & \phantom{-}\dfrac{3\sqrt{10}}{20} & -\dfrac{1}{8} 
    \end{pmatrix} 
\end{equation}
and 
\begin{equation}
    \psi^{(9)}=\begin{pmatrix}
        -\dfrac{3}{4} & \dfrac{\sqrt{10}}{2}& 1
    \end{pmatrix}^T \, .
\end{equation}
\end{enumerate}
\end{widetext}

We now move to the permutation-invariant states that have nontrivial factors of $p^{2}$ or $q^2$ in their momentum-space wave functions and are discussed in Sec.~\ref{sec:mnnonezero}. 
Listed below are the matrices $A_{i l\lambda s\,,\, jl'\lambda's'}^{(x,y)}$ and the invariant states $\psi_{i l\lambda s}$ defined in Eq.~\eqref{eqn:matAwp2q2}. 
Those matrices are spanned on $\Gamma_x$ and $\Gamma_y$, defined in Eq.~\eqref{eqn:DefHGmxGmy} and the basis vectors are given by
\begin{equation}
    pq \ket{pq\Gamma_{x}}\, , \quad p^2\ket{pq\Gamma_{y}}\, ,\quad q^2\ket{pq\Gamma_{y}}\, .
\end{equation}
In order to understand the matrix elements and the wave functions of the invariant states, the ordering of the basis vectors matters: the leftmost vector corresponds to the first row or column of $A_{i l\lambda s\,,\, jl'\lambda's'}^{(x,y)}$.
If $\Gamma_{x, y}$ has more than one set of $l\lambda s$, their sequential order is determined by where they appear from top to down in Table~\ref{tab:Subspace} and Table~\ref{tab:SubspaceII}.

\begin{widetext}
\begin{enumerate}
    \item 
    $\Gamma_x = \Gamma_{13}$, \, $\Gamma_y = \Gamma_1$ 
\begin{align}
\renewcommand{\arraystretch}{2.5} 
\mathbb{A}^{(13,1)} = 
\begin{pmatrix}
-\dfrac{1}{8} & -\dfrac{3}{8} & -\dfrac{3}{16} & -\dfrac{3}{16} & \phantom{-}\dfrac{1}{4} & \phantom{-}\dfrac{1}{4} \\
-\dfrac{3}{8} & -\dfrac{1}{8} & \phantom{-}\dfrac{3}{16} & \phantom{-}\dfrac{3}{16} & -\dfrac{1}{4} & -\dfrac{1}{4} \\
-\dfrac{3}{8} & \phantom{-}\dfrac{3}{8} & \phantom{-}\dfrac{1}{16} & -\dfrac{3}{16} & \phantom{-}\dfrac{1}{4} & -\dfrac{3}{4} \\
-\dfrac{3}{8} & \phantom{-}\dfrac{3}{8} & -\dfrac{3}{16} & \phantom{-}\dfrac{1}{16} & -\dfrac{3}{4} & \phantom{-}\dfrac{1}{4} \\
\phantom{-}\dfrac{9}{32} & -\dfrac{9}{32} & \phantom{-}\dfrac{9}{64} & -\dfrac{27}{64} & \phantom{-}\dfrac{1}{16} & -\dfrac{3}{16} \\
\phantom{-}\dfrac{9}{32} & -\dfrac{9}{32} & -\dfrac{27}{64} & \phantom{-}\dfrac{9}{64} & -\dfrac{3}{16} & \phantom{-}\dfrac{1}{16}
\end{pmatrix}
\end{align}
and
\begin{align}
\psi^{(13,1)} &= 
\begin{pmatrix}
    \phantom{-}0&\phantom{-}0&-1 & \phantom{-}1&-\dfrac{3}{4}&\phantom{-}\dfrac{3}{4}
\end{pmatrix}^T \, ,\\
\psi'^{(13,1)} &= 
\begin{pmatrix}
    -1&\phantom{-}1&\phantom{-}1 & \phantom{-}1&-\dfrac{3}{4}&-\dfrac{3}{4}
\end{pmatrix}^T\, . \label{eqn:psiprime131}
\end{align}

\item 
$\Gamma_x=\Gamma_{14}$,\, $\Gamma_y=\Gamma_{10}$

\begin{align}
    \renewcommand{\arraystretch}{2.5} 
    \mathbb{A}^{(14,10)} = \begin{pmatrix}
        \phantom{-}\dfrac{1}{4} & -\dfrac{3}{8} & \phantom{-}\dfrac{1}{2} \\
        -\dfrac{3}{4} & -\dfrac{1}{8} & -\dfrac{1}{2} \\
        \phantom{-}\dfrac{9}{16} & -\dfrac{9}{32} & -\dfrac{1}{8}
        \end{pmatrix}
\end{align}
and
\begin{align}
\psi^{(14,10)} &= 
\begin{pmatrix}
    \dfrac{4}{3} & -\dfrac{4}{3}&1
\end{pmatrix}^T \, .
\end{align}

\item $\Gamma_x=\Gamma_{15}$,\, $\Gamma_y=\Gamma_{11}$.

\begin{align}
    \renewcommand{\arraystretch}{2.5} 
    \mathbb{A}^{(15,11)} = \begin{pmatrix}
        \phantom{-}\dfrac{1}{4} & -\dfrac{3}{8} & \phantom{-}\dfrac{1}{2} \\
        -\dfrac{3}{4} & -\dfrac{1}{8} & -\dfrac{1}{2} \\
        \phantom{-}\dfrac{9}{16} & -\dfrac{9}{32} & -\dfrac{1}{8}
        \end{pmatrix}
\end{align}
and
\begin{align}
\psi^{(15,11)} &= 
\begin{pmatrix}
    \dfrac{4}{3} & -\dfrac{4}{3}&1
\end{pmatrix}^T \, .
\end{align}

\end{enumerate}

\end{widetext}
\bibliography{Refs3NF.bib}

@article{Kievsky:2016kzb,
    author = "Kievsky, A. and Viviani, M. and Gattobigio, M. and Girlanda, L.",
    title = "{Implications of Efimov physics for the description of three and four nucleons in chiral effective field theory}",
    eprint = "1610.09858",
    archivePrefix = "arXiv",
    primaryClass = "nucl-th",
    doi = "10.1103/PhysRevC.95.024001",
    journal = "Phys. Rev. C",
    volume = "95",
    number = "2",
    pages = "024001",
    year = "2017"
}

@article{Beane:2000fi,
    author = "Beane, Silas R. and Savage, Martin J.",
    title = "{Rearranging pionless effective field theory}",
    eprint = "nucl-th/0011067",
    archivePrefix = "arXiv",
    reportNumber = "NT-UW-00-028, JLAB-THY-00-63, NT@UW-00-028",
    doi = "10.1016/S0375-9474(01)01088-0",
    journal = "Nucl. Phys. A",
    volume = "694",
    pages = "511--524",
    year = "2001"
}

@article{Konig:2016utl,
    author = {K\"onig, Sebastian and Grie\ss{}hammer, Harald W. and Hammer, H. W. and van Kolck, U.},
    title = "{Nuclear Physics Around the Unitarity Limit}",
    eprint = "1607.04623",
    archivePrefix = "arXiv",
    primaryClass = "nucl-th",
    reportNumber = "INT-PUB-16-019",
    doi = "10.1103/PhysRevLett.118.202501",
    journal = "Phys. Rev. Lett.",
    volume = "118",
    number = "20",
    pages = "202501",
    year = "2017"
}

@article{Glockle:1996jg,
    author = "Glockle, Walter and Witala, H. and Huber, D. and Kamada, H. and Golak, J.",
    title = "{The Three nucleon continuum: Achievements, challenges and applications}",
    doi = "10.1016/0370-1573(95)00085-2",
    journal = "Phys. Rept.",
    volume = "274",
    pages = "107--285",
    year = "1996"
}

@article{Dietz:2021haj,
    author = {Dietz, Sebastian and Hammer, Hans-Werner and K\"onig, Sebastian and Schwenk, Achim},
    title = "{Three-body resonances in pionless effective field theory}",
    eprint = "2109.11356",
    archivePrefix = "arXiv",
    primaryClass = "nucl-th",
    doi = "10.1103/PhysRevC.105.064002",
    journal = "Phys. Rev. C",
    volume = "105",
    number = "6",
    pages = "064002",
    year = "2022"
}

@article{Weinberg:1990rz,
    author = "Weinberg, Steven",
    title = "{Nuclear forces from chiral Lagrangians}",
    reportNumber = "UTTG-31-90",
    doi = "10.1016/0370-2693(90)90938-3",
    journal = "Phys. Lett. B",
    volume = "251",
    pages = "288--292",
    year = "1990"
}

@article{Weinberg:1991um,
    author = "Weinberg, Steven",
    title = "{Effective chiral Lagrangians for nucleon - pion interactions and nuclear forces}",
    reportNumber = "UTTG-03-91",
    doi = "10.1016/0550-3213(91)90231-L",
    journal = "Nucl. Phys. B",
    volume = "363",
    pages = "3--18",
    year = "1991"
}

@article{vanKolck:1999mw,
    author = "van Kolck, U.",
    title = "{Effective field theory of nuclear forces}",
    eprint = "nucl-th/9902015",
    archivePrefix = "arXiv",
    reportNumber = "KRL-MAP-247",
    doi = "10.1016/S0146-6410(99)00097-6",
    journal = "Prog. Part. Nucl. Phys.",
    volume = "43",
    pages = "337--418",
    year = "1999"
}

@article{Kaplan:1996xu,
    author = "Kaplan, David B. and Savage, Martin J. and Wise, Mark B.",
    title = "{Nucleon - nucleon scattering from effective field theory}",
    eprint = "nucl-th/9605002",
    archivePrefix = "arXiv",
    reportNumber = "DOE-ER-40561-257, INT-96-00-125, UW-PT-96-06, CMU-HEP-96-06, DOE-ER-40862-117, CALT-68-2047",
    doi = "10.1016/0550-3213(96)00357-4",
    journal = "Nucl. Phys. B",
    volume = "478",
    pages = "629--659",
    year = "1996"
}

@book{Glockle-Fewbody83,
    author = "W. Glöckle",
    title = "{The Quantum Mechanical Few-Body Problem}",
    isbn = "978-3-642-82081-6",
    publisher = "Springer, Berlin, Heidelberg",
    month = "8",
    year = "1983"
}

@article{Hebeler:2020ocj,
    author = "Hebeler, Kai",
    title = "{Three-nucleon forces: Implementation and applications to atomic nuclei and dense matter}",
    eprint = "2002.09548",
    archivePrefix = "arXiv",
    primaryClass = "nucl-th",
    doi = "10.1016/j.physrep.2020.08.009",
    journal = "Phys. Rept.",
    volume = "890",
    pages = "1--116",
    year = "2021"
}

@article{Epelbaum:2002vt,
    author = "Epelbaum, E. and Nogga, A. and Gloeckle, Walter and Kamada, H. and Meissner, Ulf G. and Witala, H.",
    title = "{Three nucleon forces from chiral effective field theory}",
    eprint = "nucl-th/0208023",
    archivePrefix = "arXiv",
    reportNumber = "FZJ-IKP-TH-2002-18",
    doi = "10.1103/PhysRevC.66.064001",
    journal = "Phys. Rev. C",
    volume = "66",
    pages = "064001",
    year = "2002"
}

@article{Navratil:2009ut,
    author = "Navratil, Petr and Quaglioni, Sofia and Stetcu, Ionel and Barrett, Bruce R.",
    title = "{Recent developments in no-core shell-model calculations}",
    eprint = "0904.0463",
    archivePrefix = "arXiv",
    primaryClass = "nucl-th",
    reportNumber = "LLNL-JRNL-411567",
    doi = "10.1088/0954-3899/36/8/083101",
    journal = "J. Phys. G",
    volume = "36",
    pages = "083101",
    year = "2009"
}

@article{Bedaque:1997qi,
    author = "Bedaque, Paulo F. and van Kolck, U.",
    title = "{Nucleon deuteron scattering from an effective field theory}",
    eprint = "nucl-th/9710073",
    archivePrefix = "arXiv",
    reportNumber = "DOE-ER-40561-342, INT-97-00-181, DOE-ER-41014-37-N97",
    doi = "10.1016/S0370-2693(98)00430-4",
    journal = "Phys. Lett. B",
    volume = "428",
    pages = "221--226",
    year = "1998"
}

@article{vanKolck:1998bw,
    author = "van Kolck, U.",
    title = "{Effective field theory of short range forces}",
    eprint = "nucl-th/9808007",
    archivePrefix = "arXiv",
    reportNumber = "KRL-MAP-230, NT-UW-98-01",
    doi = "10.1016/S0375-9474(98)00612-5",
    journal = "Nucl. Phys. A",
    volume = "645",
    pages = "273--302",
    year = "1999"
}

@article{Bedaque:1998mb,
    author = "Bedaque, Paulo F. and Hammer, H. W. and van Kolck, U.",
    title = "{Effective theory for neutron deuteron scattering: Energy dependence}",
    eprint = "nucl-th/9802057",
    archivePrefix = "arXiv",
    reportNumber = "DOE-ER-40561-356, INT-98-00-4, TRI-PP-98-2, KRL-MAP-219",
    doi = "10.1103/PhysRevC.58.R641",
    journal = "Phys. Rev. C",
    volume = "58",
    pages = "R641--R644",
    year = "1998"
}

@article{Girlanda:2011fh,
    author = "Girlanda, L. and Kievsky, A. and Viviani, M.",
    title = "{Subleading contributions to the three-nucleon contact interaction}",
    eprint = "1102.4799",
    archivePrefix = "arXiv",
    primaryClass = "nucl-th",
    doi = "10.1103/PhysRevC.84.014001",
    journal = "Phys. Rev. C",
    volume = "84",
    number = "1",
    pages = "014001",
    year = "2011",
    note = "[Erratum: Phys.Rev.C 102, 019903 (2020)]"
}

@article{Girlanda:2020pqn,
    author = "Girlanda, L. and Kievsky, A. and Marcucci, L. E. and Viviani, M.",
    title = "{Unitary ambiguity of NN contact interactions and the 3N force}",
    eprint = "2007.04161",
    archivePrefix = "arXiv",
    primaryClass = "nucl-th",
    doi = "10.1103/PhysRevC.102.064003",
    journal = "Phys. Rev. C",
    volume = "102",
    pages = "064003",
    year = "2020"
}

@article{Nasoni:2023adf,
    author = "Nasoni, Alessia and Filandri, Elena and Girlanda, Luca",
    title = "{Relativistic constraints on 3N contact interactions}",
    eprint = "2308.13341",
    archivePrefix = "arXiv",
    primaryClass = "nucl-th",
    doi = "10.1140/epja/s10050-023-01185-3",
    journal = "Eur. Phys. J. A",
    volume = "59",
    number = "12",
    pages = "293",
    year = "2023"
}

@article{Beane:1997pk,
    author = "Beane, S. R. and Cohen, T. D. and Phillips, Daniel R.",
    title = "{The Potential of effective field theory in N N scattering}",
    eprint = "nucl-th/9709062",
    archivePrefix = "arXiv",
    reportNumber = "DOE-ER-40762-131, UMD-PP-98-024",
    doi = "10.1016/S0375-9474(98)00007-4",
    journal = "Nucl. Phys. A",
    volume = "632",
    pages = "445--469",
    year = "1998"
}

@article{Cohen:1996my,
    author = "Cohen, Thomas D.",
    title = "{Regularization, renormalization and range: The Nucleon-nucleon interaction from effective field theory}",
    eprint = "nucl-th/9606044",
    archivePrefix = "arXiv",
    reportNumber = "DOE-ER-40762-087, UMD-PP-96-112",
    doi = "10.1103/PhysRevC.55.67",
    journal = "Phys. Rev. C",
    volume = "55",
    pages = "67--72",
    year = "1997"
}

@article{Scaldeferri:1996nx,
    author = "Scaldeferri, K. A. and Phillips, Daniel R. and Kao, C. W. and Cohen, T. D.",
    title = "{Short range interactions in an effective field theory approach for nucleon-nucleon scattering}",
    eprint = "nucl-th/9610049",
    archivePrefix = "arXiv",
    reportNumber = "UMD-PP-97-053, DOE-ER-40762-105",
    doi = "10.1103/PhysRevC.56.679",
    journal = "Phys. Rev. C",
    volume = "56",
    pages = "679--688",
    year = "1997"
}

@article{Gegelia:1998gn,
    author = "Gegelia, J.",
    title = "{EFT and N N scattering}",
    doi = "10.1016/S0370-2693(98)00460-2",
    journal = "Phys. Lett. B",
    volume = "429",
    pages = "227--231",
    year = "1998"
}

@article{Barrett:2004qsq,
    author = "Barrett, B. R. and Navr\'atil, P. and Nogga, A. and Ormand, W. E. and Vary, J. P.",
    editor = "Savard, G. and Davids, C. N. and Lister, C. J.",
    title = "{No-core shell-model calculations in light nuclei with three-nucleon forces}",
    doi = "10.1016/j.nuclphysa.2004.09.137",
    journal = "Nucl. Phys. A",
    volume = "746",
    pages = "579--582",
    year = "2004"
}

@article{Konig:2019xxk,
    author = {K\"onig, Sebastian},
    title = "{Energies and radii of light nuclei around unitarity}",
    eprint = "1910.12627",
    archivePrefix = "arXiv",
    primaryClass = "nucl-th",
    doi = "10.1140/epja/s10050-020-00098-9",
    journal = "Eur. Phys. J. A",
    volume = "56",
    number = "4",
    pages = "113",
    year = "2020"
}

@article{Stetcu:2006ey,
    author = "Stetcu, I. and Barrett, B. R. and van Kolck, U.",
    title = "{No-core shell model in an effective-field-theory framework}",
    eprint = "nucl-th/0609023",
    archivePrefix = "arXiv",
    doi = "10.1016/j.physletb.2007.07.065",
    journal = "Phys. Lett. B",
    volume = "653",
    pages = "358--362",
    year = "2007"
}

@article{Hammer:2019poc,
    author = {Hammer, H. -W. and K\"onig, S. and van Kolck, U.},
    title = "{Nuclear effective field theory: status and perspectives}",
    eprint = "1906.12122",
    archivePrefix = "arXiv",
    primaryClass = "nucl-th",
    doi = "10.1103/RevModPhys.92.025004",
    journal = "Rev. Mod. Phys.",
    volume = "92",
    number = "2",
    pages = "025004",
    year = "2020"
}

@article{vanKolck:1994yi,
    author = "van Kolck, U.",
    title = "{Few nucleon forces from chiral Lagrangians}",
    doi = "10.1103/PhysRevC.49.2932",
    journal = "Phys. Rev. C",
    volume = "49",
    pages = "2932--2941",
    year = "1994"
}

@article{Bernard:2007sp,
    author = "Bernard, V. and Epelbaum, E. and Krebs, H. and Meissner, Ulf-G.",
    title = "{Subleading contributions to the chiral three-nucleon force. I. Long-range terms}",
    eprint = "0712.1967",
    archivePrefix = "arXiv",
    primaryClass = "nucl-th",
    reportNumber = "FZJ-IKP-TH-2007-33, HISKP-TH-07-27",
    doi = "10.1103/PhysRevC.77.064004",
    journal = "Phys. Rev. C",
    volume = "77",
    pages = "064004",
    year = "2008"
}

@article{Bernard:2011zr,
    author = "Bernard, V. and Epelbaum, E. and Krebs, H. and Meissner, U. -G.",
    title = "{Subleading contributions to the chiral three-nucleon force II: Short-range terms and relativistic corrections}",
    eprint = "1108.3816",
    archivePrefix = "arXiv",
    primaryClass = "nucl-th",
    doi = "10.1103/PhysRevC.84.054001",
    journal = "Phys. Rev. C",
    volume = "84",
    pages = "054001",
    year = "2011"
}

@article{Krebs:2012yv,
    author = "Krebs, Hermann and Gasparyan, A. and Epelbaum, Evgeny",
    title = "{Chiral three-nucleon force at N$^4$LO I: Longest-range contributions}",
    eprint = "1203.0067",
    archivePrefix = "arXiv",
    primaryClass = "nucl-th",
    doi = "10.1103/PhysRevC.85.054006",
    journal = "Phys. Rev. C",
    volume = "85",
    pages = "054006",
    year = "2012"
}

@article{Krebs:2013kha,
    author = "Krebs, Hermann and Gasparyan, A. and Epelbaum, Evgeny",
    title = "{Chiral three-nucleon force at $N^4LO$ II: Intermediate-range contributions}",
    eprint = "1302.2872",
    archivePrefix = "arXiv",
    primaryClass = "nucl-th",
    doi = "10.1103/PhysRevC.87.054007",
    journal = "Phys. Rev. C",
    volume = "87",
    number = "5",
    pages = "054007",
    year = "2013"
}

@article{Nogga:2000vq,
    author = "Nogga, A. and Kamada, H. and Gloeckle, Walter",
    editor = "Stadler, A. and Arriaga, A. and Cravo, E. and Fonseca, A. C. and Nunes, F. M. and Pena, M. T. and Rupp, G.",
    title = "{Solution of the Faddeev-Yakubovsky equations using realistic N N and 3 N interaction}",
    eprint = "nucl-th/0010005",
    archivePrefix = "arXiv",
    doi = "10.1016/S0375-9474(01)00854-5",
    journal = "Nucl. Phys. A",
    volume = "689",
    pages = "357--360",
    year = "2001"
}

@article{Navratil:2007we,
    author = "Navratil, P. and Gueorguiev, V. G. and Vary, J. P. and Ormand, W. E. and Nogga, A.",
    title = "{Structure of A=10-13 nuclei with two plus three-nucleon interactions from chiral effective field theory}",
    eprint = "nucl-th/0701038",
    archivePrefix = "arXiv",
    reportNumber = "UCRL-JRNL-227256",
    doi = "10.1103/PhysRevLett.99.042501",
    journal = "Phys. Rev. Lett.",
    volume = "99",
    pages = "042501",
    year = "2007"
}

@article{Ordonez:1993tn,
    author = "Ordonez, C. and Ray, L. and van Kolck, U.",
    title = "{Nucleon-nucleon potential from an effective chiral Lagrangian}",
    reportNumber = "UTTG-31-93, DOE-ER-40427-23-N93",
    doi = "10.1103/PhysRevLett.72.1982",
    journal = "Phys. Rev. Lett.",
    volume = "72",
    pages = "1982--1985",
    year = "1994"
}

@article{Witala:2022rzl,
    author = "Wita{\l}a, H. and Golak, J. and Skibi{\'n}ski, R.",
    title = "{Significance of chiral three-nucleon force contact terms for understanding of elastic nucleon-deuteron scattering}",
    eprint = "2203.08499",
    archivePrefix = "arXiv",
    primaryClass = "nucl-th",
    doi = "10.1103/PhysRevC.105.054004",
    journal = "Phys. Rev. C",
    volume = "105",
    number = "5",
    pages = "054004",
    year = "2022"
}

\end{document}